\documentclass{ieeeaccess}
\usepackage{cite}
\usepackage{amsmath,amssymb,amsfonts}
\usepackage{algorithmic}
\usepackage{graphicx}
\usepackage{textcomp}

\usepackage{textcomp}
\usepackage[utf8]{inputenc}
\usepackage{textgreek}
\usepackage{caption}
\usepackage{multirow}
\usepackage[hyphens]{url}
\usepackage{hyperref}
\usepackage{xcolor}

\def\BibTeX{{\rm B\kern-.05em{\sc i\kern-.025em b}\kern-.08em
    T\kern-.1667em\lower.7ex\hbox{E}\kern-.125emX}}
\begin{document}

\history{Date of publication xxxx 00, 0000, date of current version xxxx 00, 0000.}
\doi{10.1109/ACCESS.2017.DOI}

\title{PowerModels-ACOPF-AI: On-the-Fly Machine Learning Approach for Solving AC Optimal Power Flow Integrating Renewable Energy Sources}
\author {\uppercase{Bhuban Dhamala}\authorrefmark{1}, \IEEEmembership{Student Member, IEEE},
\uppercase{Jose E. Tabarez\authorrefmark{2},
\uppercase{Anup Pandey}\authorrefmark{3}}}
\address[1]{The University of Texas at Dallas, Richardson, TX 75080 USA (e-mail: bhuban.dhamala@utdallas.edu )}
\address[2]{Los Alamos National Laboratory, Los Alamos, NM, 87545 USA}
\address[3]{Los Alamos National Laboratory, Los Alamos, NM, 87545 USA (e-mail: anup@lanl.gov)}
\tfootnote{Research presented in this paper was supported by the Laboratory Directed Research and Development program of Los Alamos National Laboratory under project number 20250854ECR.}

\markboth
{Author \headeretal: Preparation of Papers for IEEE TRANSACTIONS and JOURNALS}
{Author \headeretal: Preparation of Papers for IEEE TRANSACTIONS and JOURNALS}

\corresp{Corresponding author: Bhuban Dhamala (e-mail: bhuban.dhamala@utdallas.edu)}

\begin{abstract}
The increasing complexity of modern power systems, driven by high renewable penetration, load variability, and operational uncertainty, demands fast and reliable solutions to the AC optimal power flow problem (AC-OPF). Traditional optimization methods, though accurate, often struggle with scalability and high computational burdens, making them impractical for real-time use in large networks. This paper introduces PowerModels-ACOPF-AI, a two-stage Bayesian Neural Network (BNN) surrogate designed to predict generator set points, bus voltages, and phase angles with uncertainty. The framework integrates a performance-sensitive on-the-fly learning mechanism that identifies regions of degraded prediction accuracy and dynamically retrains with additional AC-OPF solutions generated by PowerModels.jl. This self-adaptive loop ensures robust performance, enabling the model to maintain accuracy under novel or highly variable operating conditions. Validation on benchmark test systems of different sizes—namely the 30-bus, 200-bus, and 500-bus networks—demonstrates strong generalization capability, efficient handling of stochastic renewable injections, and the ability to provide rapid, uncertainty-aware predictions. Beyond predictive accuracy, the proposed approach offers practical value as both a real-time advisory tool for system operators and a fast initializer for conventional solvers, thus supporting resilient and efficient grid operation in future power systems.
\end{abstract}

\begin{keywords}
AC Optimal Power Flow (AC-OPF), Bayesian Neural Network (BNN), Machine Learning for Power Systems, On-the-Fly learning, PowerModel-AI-OPF, PowerModels.jl, Renewable Energy Integration
\end{keywords}

\titlepgskip=-15pt

\maketitle

\section{Introduction}
\label{sec:introduction}
\PARstart{T}{he} Optimal Power Flow (OPF) serves as a foundational optimization framework in modern power system operations, aiming to determine the most efficient, secure, and reliable operating point of a power network under steady-state conditions. It optimally adjusts controllable variables—such as real and reactive generator outputs, voltage magnitudes at generator buses, transformer tap positions, and settings of reactive compensation devices—to minimize or maximize a specified objective function. Common objectives include generation cost, system losses, emission levels, and voltage deviation. The OPF formulation rigorously satisfies the non-linear power flow equations, along with a diverse set of operational constraints. These constraints include equality conditions derived from Kirchhoff’s laws and inequality limits on voltage magnitudes, power injections, line thermal ratings, and generator operating boundaries.

Originating from the classical economic dispatch formulation, the OPF extends this by embedding the full AC power flow model, which introduces nonlinearity and nonconvexity into the solution space. As a result, solving OPF necessitates the use of advanced mathematical techniques, including interior point methods, sequential quadratic programming, and, more recently, convex relaxations and metaheuristic algorithms to ensure tractable and globally optimal or near-optimal solutions. With the advent of smart grids, distributed generation, and high penetration of renewables, OPF has evolved into broader formulations, such as stochastic OPF, robust OPF, and security-constrained OPF, which are capable of handling uncertainty, contingency scenarios, and dynamic constraints. Consequently, OPF is indispensable not only for dispatch scheduling and voltage control, but also for market clearing, congestion management, and grid security assessment. Its multiscale applicability, from day-ahead planning to minute-level operational decisions, makes it a pivotal tool in both centralized control frameworks and decentralized grid architectures, underpinning the transition to resilient and sustainable power systems.

Optimal Power Flow (OPF) has long served as a foundational tool in the operational planning of power systems \cite{b4}, facilitating unit commitment \cite{b5}, security constraint economic dispatch \cite{b6}, congestion management \cite{b7}, and voltage regulation \cite{b8}, while ensuring compliance with physical and security constraints. In conventional grids dominated by dispatchable thermal units, day-ahead OPF formulations have proven sufficient for forecasting and scheduling generation in alignment with predicted demand. Grid operators typically solve the OPF problem at intervals of about fifteen minutes to ensure the power network remains secure and operates steadily \cite{b9}. However, the increasing penetration of highly intermittent renewable energy resources, such as solar photovoltaics and wind power, has introduced substantial uncertainty and variability into power system operations, fundamentally challenging the adequacy of traditional day-ahead planning frameworks. Unlike conventional generators, renewable units are nondispatchable, weather dependent, and spatially dispersed, exhibiting fast-changing generation profiles that frequently deviate from forecasts. As a result, the timescale of operational decision-making must shift from hours to minutes or even seconds, necessitating the application of OPF in real-time or near-real-time contexts. This evolution complicates the OPF landscape by introducing time-varying and stochastic power injections. These uncertainties require the use of advanced formulations such as stochastic and robust AC-OPF, which significantly increase both the dimensionality and computational burden. Consequently, implementing full AC-OPF for real-time decision making becomes extremely challenging and often impractical without substantial simplification. 

In practice, most OPF problems are solved using the linearized DC power flow approximation \cite{b2} \cite{b3}. The results of DC-OPF are then typically used to perform contingency analysis using AC power flow or to provide an initial guess to solve the nonlinear AC-OPF problem. However, the DC approximation has notable limitations, as it is inherently incapable of modeling reactive power flows and accurate bus voltage profiles, and this inefficiency by approximation may cost millions of dollars in power system operation \cite{b1}. Therefore, the pursuit of accurate, reliable, and computationally efficient AC-OPF solutions remains a critical priority in modern power system operation.

The increasing complexity and time sensitivity of modern power system operations, particularly under high penetration of renewable energy sources and unpredictably changing demand capacity, have catalyzed the exploration of machine learning (ML) techniques as promising enablers for real-time optimal power flow (OPF) solutions. Traditional optimization algorithms, while robust and accurate, often suffer from high computational burdens, especially in large systems and a wide range of generation and demand variations. Machine learning methods, by using extensive historical data or simulated OPF solutions, shift the computational burden from online optimization to offline training processes. These approaches capture complex relationships effectively and allow for rapid prediction of variables. Once trained, the ML model can learn complex input-output mappings between different states of load and generation and the optimal setpoints, such as generator output, bus voltages, power flow, etc., allowing for near-instantaneous inference once deployed. 

ML applications are categorized into several key areas: direct mapping of OPF variables, prediction of active constraints, mapping binary decision variables (for example, for unit commitment), learning control policies, incorporating stability constraints, and predicting warm-start points for faster solver convergence \cite{b10} \cite{b11}. In OPF problems, machine learning uses historical or simulated data to train models offline, with the aim of either predicting solutions quickly or speeding up traditional OPF solvers. These methods fall into two main categories: End-to-End (E2E) learning and Learning-to-Optimize (L2O). E2E directly maps inputs to OPF outputs, bypassing the optimization process. In contrast, L2O helps traditional solvers by guiding or improving parts of the optimization steps \cite{b12}. 

Recent advances have explored machine learning (ML)-based methodologies as a promising avenue for enhancing the computational efficiency of solving the AC Optimal Power Flow (AC-OPF) problem. In \cite{b14}, a neural network-based framework is introduced to emulate the iterative process of the MIPS AC-OPF solver. Rather than predicting the final solution directly, the model learns a sequence of updates that gradually steer the state of the system toward convergence. In a related effort, \cite{b15} presents an ML-based framework that integrates an Extreme Learning Machine (ELM) with power flow equations to predict bus voltage magnitudes and angles, from which power injections are computed using the admittance matrix, thus preserving physical feasibility while improving computational speed. A deep reinforcement learning (DRL) approach is proposed in \cite{b16}, where the AC-OPF is formulated as a Markov Decision Process (MDP), and control policies are learned through interactions with a simulated grid environment, using generator outputs and voltage settings as actions. To eliminate the dependency on labeled data, \cite{b17} introduces an unsupervised learning framework that trains a deep neural network to infer bus voltages from load inputs; however, the absence of labels complicates model evaluation, particularly when multiple valid solutions exist. A regionally decomposable two-stage learning architecture is proposed in \cite{b18}, where system-wide coupling variables are first predicted, followed by the training of region-specific neural networks. While this design improves scalability, it is sensitive to the choice of partitioning and may suffer from error propagation across stages. The  \cite{b19} presents a deep learning framework combining pseudo-labeling through ridge regression with a variable splitting scheme and physics-informed gradient estimation. Although effective, its reliance on gradient approximations and pseudo-label quality can limit robustness under highly variable system conditions.

In our prior work \cite{b24}, an on-the-fly learning framework was developed for fast prediction of AC power flow (AC-PF) solutions, focusing primarily on estimating feasible voltage magnitudes and phase angles without incorporating optimization objectives. In contrast, the present work significantly extends this concept to the AC Optimal Power Flow (AC-OPF) problem, which involves nonlinear optimization of generator dispatch under operational and economic constraints. Specifically, this paper introduces a two-stage Bayesian Neural Network (BNN) architecture for predicting generator setpoints and voltage profiles, incorporates uncertainty quantification in OPF solutions, explicitly models renewable generation variability, and develops a performance-aware on-the-fly retraining mechanism tailored for optimization outputs. The proposed model is validated on several systems, such as the IEEE 30-bus system, the synthetic 200-bus system, and the synthetic 500-bus network, and is capable of accurately handling complex operating conditions characterized by high penetration of non-dispatchable renewable energy sources and diverse nodal load variations. The ML workflow is executed in two sequential stages: Stage 1 estimates the active and reactive power setpoints for controllable generator units, while Stage 2 predicts the bus voltage magnitudes and phase angles across all buses. To ensure sustained predictive accuracy and adaptability, the model integrates a performance-aware learning mechanism. Specifically, it employs a criterion-based evaluation strategy to identify regions of sub-optimal prediction. In these regions, an on-the-fly learning approach is implemented, allowing the model to autonomously generate new training data through PowerModels.jl \cite{b13}, retrain itself, and dynamically update its parameters. This continuous learning loop significantly enhances the robustness, generalizability, and accuracy of the model in rapidly changing grid conditions.

\section{AC Optimal Power Flow}
The OPF problem aims to minimize costs by optimally allocating generator resources, while meeting power demands and adhering to various technical and physical constraints of the power network. In its simplified form, the OPF problem can be expressed as:

\begin{flalign}
&\min_{\{P_{Gi}\}_{i\in G}} \sum_{i\in G} C_i\!\bigl(P_{Gi}\bigr) && \tag{1}\label{eq:obj}\\[1pt]
&\text{s.t.}\; P_i(V,\theta)=P_{Gi}-P_{Li},\quad \forall i\in N && \tag{2}\label{eq:PbalR}\\
&\phantom{\text{s.t.}\;} P_i(V,\theta)=|V_i|\sum_{j=1}^{N_B}|V_j|\bigl(G_{ij}\cos\delta_{ij}+B_{ij}\sin\delta_{ij}\bigr) && \tag{2a}\label{eq:Pflow}\\
&\phantom{\text{s.t.}\;} Q_i(V,\theta)=Q_{Gi}-Q_{L_i}, \quad \forall i\in N && \tag{3}\label{eq:QbalR}\\
&\phantom{\text{s.t.}\;} Q_i(V,\theta)=|V_i|\sum_{j=1}^{N_B}|V_j| \bigl(G_{ij}\sin\delta_{ij}-B_{ij}\cos\delta_{ij}\bigr) && \tag{3a}\label{eq:Qflow}\\
&\phantom{\text{s.t.}\;} P_{G_i}^{\min}\le P_{G_i}\le P_{G_i}^{\max}, \quad \forall i\in G && \tag{4}\label{eq:Plim}\\
&\phantom{\text{s.t.}\;} Q_{G_i}^{\min}\le Q_{G_i}\le Q_{G_i}^{\max}, \quad \forall i\in G && \tag{5}\label{eq:Qlim}\\
&\phantom{\text{s.t.}\;} V_i^{\min}\le |V_i|\le V_i^{\max}, \quad \forall i\in N  && \tag{6}\label{eq:Vlim}\\
&\phantom{\text{s.t.}\;} \theta_i^{\min}\le \theta_i\le \theta_i^{\max}, \quad \forall i\in N && \tag{7}\label{eq:anglim}
\end{flalign}

The objective function~\eqref{eq:obj} minimizes the total production cost by summing the individual generation cost curves \(C_i(P_{G_i})\) over all controllable units \(i\in G\).
The power balance constraints~\eqref{eq:PbalR} and~\eqref{eq:QbalR} enforce the Kirchhoff current law on every bus \(i\in N\), comparing the scheduled real and reactive injections \((P_{G_i},Q_{G_i})\) with the corresponding demands \((P_{L_i},Q_{L_i})\) and the calculated injections in the network \(P_i(V,\theta)\) and \(Q_i(V,\theta)\). These injections are given by the nonlinear AC power-flow relations~\eqref{eq:Pflow} and~\eqref{eq:Qflow}, which couple every pair of buses through the bus-admittance coefficients \((G_{ij},B_{ij})\) and the voltage variables \(|V_i|\) and \(\theta_i\).Therefore, the OPF solution satisfies the physical requirement that total generation equals total demand plus network losses. Thus, both local (bus-level) and global (system-level) power conservation are inherently maintained in any feasible AC-OPF solution. Generator capability curves are enforced by the active and reactive limits~\eqref{eq:Plim}–\eqref{eq:Qlim}, while operational security is preserved by maintaining voltage magnitudes and phase angles within the bounds~\eqref{eq:Vlim}–\eqref{eq:anglim}. Together, these constraints yield a non-convex optimization problem whose solution provides an economically efficient and electrically feasible dispatch that complies with system-reliability standards.

%
\section{Bayesian Neural Network}
Neural networks have emerged as powerful tools for modeling complex, high-dimensional functions across various domains, including power systems. However, conventional neural networks produce point estimates of model parameters—fixed weights and biases—which makes them incapable of capturing uncertainty in predictions. This limitation poses a significant challenge in applications that require robust decision-making under uncertainty. Bayesian Neural Networks (BNNs) address this issue by extending traditional architectures through the incorporation of Bayesian inference. Instead of assigning fixed values to weights and biases, BNNs treat them as probability distributions, allowing the model to quantify uncertainty in its predictions. This probabilistic framework enables the integration of prior knowledge and facilitates the continuous updating of parameter beliefs as new data becomes available, leading to more reliable and informed predictions.

Bayesian inference is a statistical paradigm that updates the probability distribution of the parameters of a model based on the observed data. Given a data set \( D \) and model parameters \( \theta \), Bayesian inference computes the posterior distribution of the parameters using the Bayes theorem \cite{b21}:

\begin{align}
    p(\theta \mid D) = \frac{p(D \mid \theta) \cdot p(\theta)}{p(D)} \tag{8} \label{eq:Bayes' Theorem}
\end{align}

Here, \( p(\theta) \) is the prior distribution, \( p(D \mid \theta) \) is the likelihood, and \( p(D) \) is the marginal likelihood or evidence. The posterior \( p(\theta \mid D) \) reflects the updated beliefs after observing the data.

In Bayesian neural networks (BNNs), this framework is applied by treating network weights and biases as probability distributions rather than fixed values. Consequently, the output becomes a predictive distribution obtained by integrating over the posterior:

\begin{align}
  p(y \mid x, D) = \int p(y \mid x, \theta) \, p(\theta \mid D) \, d\theta \tag{9}\label{eq:BNN1}
\end{align}

This approach enables BNNs to quantify uncertainty in predictions, offering improved robustness and interpretability in data-driven modeling.
To make predictions for a new input \( x^* \), we compute the posterior predictive distribution:

\begin{align}
p(y^* \mid x^*, D) = \int p(y^* \mid x^*, \theta) \, p(\theta \mid D) \, d\theta \tag{10}
\end{align}\label{eq:BNN2}

where \( y^* \) is the predicted output. This integration accounts for uncertainty in the model parameters, which yields more robust and calibrated predictions.

The on-the-fly learning mechanism refers to a dynamic, real-time model adaptation strategy in which the machine learning system continuously monitors its prediction performance and autonomously triggers updates when accuracy deteriorates. Rather than requiring a full retraining with a vast dataset upfront, the model assesses specific criteria to determine if retraining is necessary. When any of these specified conditions are unmet, the system automatically generates new data in the problematic region, re-trains itself using this data, and updates its parameters accordingly. This adaptive process ensures sustained accuracy and reliability of predictions in power systems across varying grid conditions without interrupting system operations.

BNNs can enhance AC Optimal Power Flow (AC-OPF) studies by offering an uncertainty-aware surrogate that complements conventional optimization solvers. Trained on historical or simulated data, a BNN learns the nonlinear mapping from system inputs—load levels, renewable generation profiles, network topologies— to OPF outputs such as generator set-points, bus voltages, and branch flows. Because renewable sources (e.g., wind) are inherently intermittent, non-dispatchable, and difficult to forecast on short time scales, modeling their injections as stochastic inputs and training the BNN on this randomness enables the network to predict optimal control actions for dispatchable units while explicitly capturing operational uncertainty. The resulting surrogate provides rapid and probabilistic recommendations that support real-time decision making and robust dispatch under volatile operating conditions.

A common approach to generate synthetic wind power outputs is to sample either from a uniform distribution or a truncated normal distribution confined within the physical limits of the plant, as expressed in \eqref{eq:Pwind_dist}; the rated capacity \(P_{\text{rated}}\) enforces the hard limits \(0 \le P_{\text{res}} \le P_{\text{rated}}\) at feasible production levels.

\begin{equation}
P_{\text{wind}} \sim
\begin{cases}
\mathcal{U}\!\bigl(0,\,P_{\text{rated}}\bigr), & \text{(uniform model)},\\[6pt]
\mathcal{N}\!\bigl(\mu,\,\sigma^{2}\bigr)\,\bigl|\,[0,\,P_{\text{rated}}], & \text{(truncated normal model)}\tag{11}\label{eq:Pwind_dist}
\end{cases}
\end{equation}

For each realization of wind generation \(P_{\text{res}}^{(i)}\), an AC-OPF problem is solved with a conventional nonlinear solver (e.g., IPOPT), giving optimal set-points \(P_G^{(i)}, Q_G^{(i)}, V^{(i)}, \theta^{(i)}\).  The AC-OPF is cast as  
\[
\begin{array}{ll}
\displaystyle \min_{\{P_{G_i}\}_{i\in\mathcal{G}}} & \displaystyle \sum_{i\in\mathcal{G}} C_i\!\bigl(P_{G_i}\bigr) \\[8pt]
\text{subject to} &
\begin{cases}
P_i(V,\theta) = P_{G_i}+P_{\text{res},i} -P_{L_i}, \\[4pt]
Q_i(V,\theta) = Q_{G_i}+Q_{\text{res},i}-Q_{L_i}, \\[4pt]
Eqs. (4) - (7)\tag{12}\label{eq:OPF_with_wind}
\end{cases}
\end{array}
\]
where \(x=[P_G,Q_G,V,\theta]\) is the decision vector, \(C_i(P_{G_i})\) the cost of the generator \(i\), and \(\mathcal{G}\) the sets of generators.  The input vector \(\mathbf{u}^{(i)}=[P_{\text{load}}^{(i)}, Q_{\text{load}}^{(i)}, P_{\text{res}}^{(i)}]\) together with the optimal solution \(x^{(i)}\) forms the training set \(\{\mathbf{u}^{(i)}, x^{(i)}\}_{i=1}^{N}\); a Bayesian neural network is trained in this set to learn the mapping \(\mathbf{u}\mapsto x\) while quantifying predictive uncertainty. This probabilistic framework enables the BNN to provide not only point estimates of generator setpoints and voltage profiles, but also credible intervals that reflect the uncertainty arising from changes in loads and variability of renewable resources. 

\subsection{Two Stage Architecture for AC-OPF}
This AC-OPF surrogate framework, utilizing Bayesian Neural Networks (BNNs) is implemented in a two-stage architecture designed to emulate the sequential decision process of conventional OPF solvers while capturing model uncertainty.
\subsubsection{Stage 1: Generation Set Points}
In the first stage, a BNN model is constructed to approximate the optimal active and reactive power set points of controllable generation units, as well as the reactive power set points of renewable energy resources. The latter is particularly important for compliance with the federal Energy Regulatory Commission (FERC) interconnection requirements for inverter-based resources (IBRs) [22]. 
The input vector to this stage is defined as:
\[
\mathbf{u}_1^{(i)} = \left[P_{\text{load}}^{(i)}, P_{\text{res}}^{(i)}, Q_{\text{load}}^{(i)}\right]
\]
The corresponding output vector from the first-stage BNN is:
\[
\mathbf{x}_1^{(i)} = \left[P_G^{(i)}, Q_G^{(i)}, Q_{\text{res}}^{(i)}\right]
\]

This output comprises the active and reactive power dispatch values of the generator alongside the reactive setpoints of renewable sources.

\vspace{0.5em}

\subsubsection{Stage 2: Voltage Profile}
Upon obtaining the optimal dispatch of the first-stage generation from the first-stage model, a second-stage BNN is trained to predict the voltage profile of the power system, specifically the bus voltage magnitudes \( V \) and phase angles \( \theta \). The motivation for this decoupling is twofold: first, to reduce the input dimensionality and target complexity of each surrogate model, thereby enhancing the learning efficiency and prediction accuracy; second, to reflect the inherent hierarchical structure of traditional OPF solution procedures, where generation dispatch is typically solved before determining the steady-state system voltages.
The input vector to the second-stage BNN is defined as:
\[
\mathbf{u}_2^{(i)} = \left[P_{\text{load}}^{(i)}, Q_{\text{load}}^{(i)}, P_G^{(i)}, Q_G^{(i)}, P_{\text{res}}^{(i)}, Q_{\text{res}}^{(i)}\right]
\]

The corresponding output vector is:
\[
\mathbf{x}_2^{(i)} = \left[V^{(i)}, \theta^{(i)}\right]
\]

This output represents the steady-state operating point of the system.

Figure 1 illustrates the general workflow to perform an optimal AC load flow analysis using the proposed approach. The process begins with the execution of AC optimal power flow (OPF) simulations and the collection of the corresponding raw data. This is followed by a data analysis phase, in which the quality of the data is evaluated and essential patterns are identified. In the subsequent Feature Selection step, relevant input variables (features) and corresponding outputs (targets) are selected to train the ML model. These selected features are then utilized in the Model Training phase to develop a BNN model capable of estimating OPF results. The trained model undergoes an on-the-fly implementaton. This step actively identifies regions of high prediction error, generates additional data in those regions, and retrains the model accordingly, thus improving its accuracy and generalization performance.

Figure 2 presents the overall architecture of the proposed on-the-fly AC-OPF framework as a sequential, yet adaptive process. As illustrated earlier, the training
procedure is executed in two stages. Key steps in the process include generating and formatting AC OPF datasets, selecting appropriate features and targets of the model, and implementing the on-the-fly validation mechanism for continuous model refinement. The first stage comprised of four key steps: initialization, real-time inference, performance monitoring, and adaptive retraining. The process begins with an initialization stage, where a Bayesian Neural Network (BNN) is trained using a representative dataset of AC-OPF solutions generated from conventional solvers. This trained model serves as the starting point for the operation. Once deployed, the framework enters the inference stage, where incoming system conditions—such as load variations and renewable generation uncertainty—are fed into the BNN. The model produces not only predicted AC-OPF solutions but also associated uncertainty estimates, which are critical for assessing prediction reliability. The next step in the flowchart is continuous performance monitoring, where the framework evaluates whether the model remains reliable under current operating conditions. This is achieved by checking if the true system’s behavior remains statistically consistent with the predicted uncertainty bounds. As long as the model maintains acceptable reliability, the system continues operating in this fast inference mode.

However, if the monitoring block detects degraded performance, for example, due to unseen operating conditions or high uncertainty, the framework activates the adaptive retraining loop. In this stage, the system selectively generates new data by running targeted AC-OPF simulations in the regions where the model is under performing. These newly generated samples are then used to update the BNN. Importantly, this update is incremental and avoids retraining on the entire dataset, making the process computationally efficient. Finally, the updated model is fed back into the inference stage, completing the closed-loop structure shown in the flowchart. This cyclical process ensures that the model continuously improves over time while maintaining real-time applicability. 
The second stage involves developing a BNN model for a detailed analysis of the AC load flow analysis to predict the each bus voltage magnitude and bus voltage angle. The specific inputs and outputs used in each stage are shown in Figure 2.

The proposed framework monitors predictive reliability using the uncertainty quantification capability of the Bayesian Neural Network (BNN), specifically through the ±2σ confidence interval. Model performance is evaluated based on statistical coverage, defined as the proportion of true AC-OPF solutions contained within the predicted uncertainty band. A degradation in accuracy is identified when this coverage falls below a predefined threshold (e.g., 90\%), indicating operation in out-of-distribution conditions. In such cases, a simulation-driven data generation process is triggered using PowerModels.jl, where new operating points are sampled in the identified region, and corresponding AC-OPF solutions are computed using a conventional nonlinear solver. These samples are incorporated to retrain the model while preserving its original features, thereby restoring prediction accuracy.

\begin{figure}[!t]
  \centering
  \includegraphics[width=0.6\linewidth]{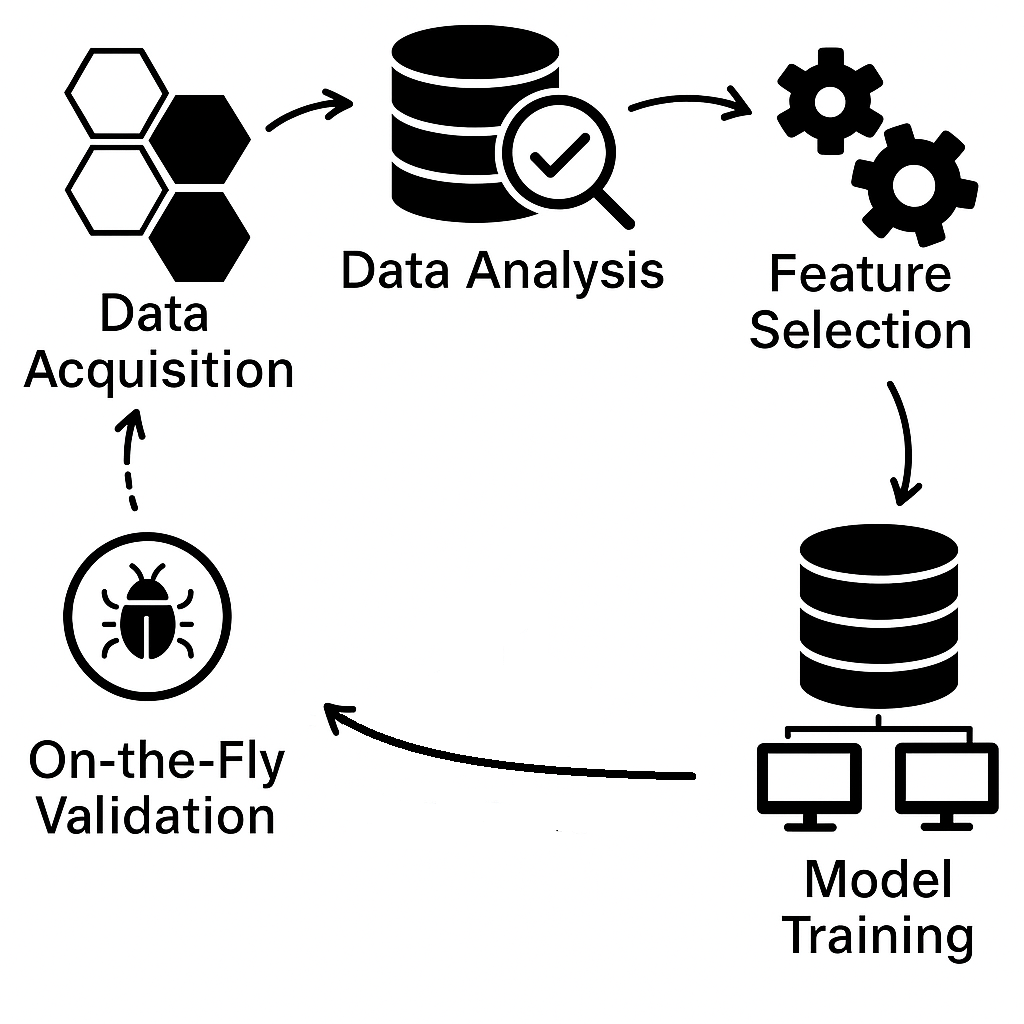}
  \caption{\textbf{Overview of the PowerModels-ACOPF-AI}}
  \label{fig2}
\end{figure}

\begin{figure}[!t]
  \centering
  \includegraphics[width=1\linewidth]{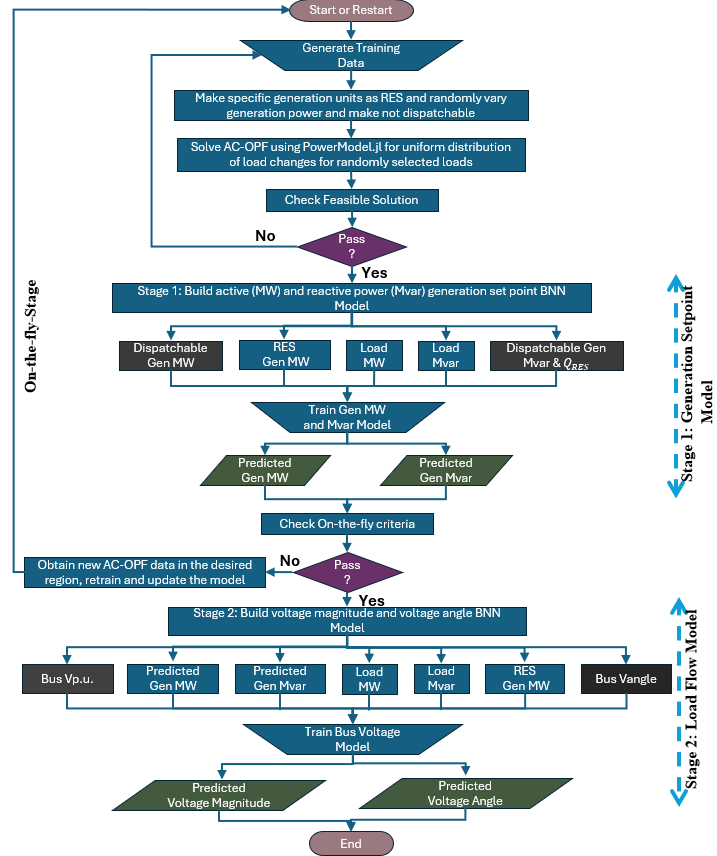}

  \caption{\textbf{Schematic representation of PowerModels-ACOPF-AI for optimal power flow with renewable integration.}}
  \label{fig: Flow Chart}
\end{figure}

\section{Training Dataset}
All training datasets used in this study, which serve as the ground truth for model development and validation, were generated using PowerModels.jl, an open-source optimization framework developed in the Julia programming language specifically for power system analysis \cite{b13}. Built on top of the JuMP modeling language, PowerModels.jl provides a flexible, extensible, and high-performance environment for formulating and solving a wide range of power system optimization problems, including AC Optimal Power Flow (AC-OPF). The framework seamlessly integrates with state-of-the-art solvers such as IPOPT, enabling the application of advanced nonlinear and iterative optimization techniques to large-scale power networks. 

To evaluate and validate the proposed PowerModels-ACOPF-AI framework, extensive analyses were conducted on the standard IEEE 30-bus system, a 200-bus synthetic test system, and a 500-bus synthetic test system \cite{b20}. The IEEE 30-bus system consists of 6 generator units and 20 load buses, whereas the 200-bus and 500-bus synthetic systems comprise 49 and 60 generator buses and 108 and 200 load buses, respectively, representing progressively larger and more complex network topologies.
A total of 7,623 feasible AC Optimal Power Flow (AC-OPF) solutions were obtained for the IEEE 30-bus system, 7,991 for the 200-bus system, and 9,978 for the 500-bus system. These correspond to the subsets of randomly generated input scenarios for which the OPF problem successfully converged. Specifically, the 7,623 feasible solutions were obtained from 12,000 random input instances for the 30-bus system, 7,991 from 10,000 for the 200-bus system, and 9,978 from 10,000 for the 500-bus system.
For each instance, a subset of load buses was randomly selected, and their active and reactive power demands were modified by scaling the nominal load values using random factors drawn from a uniform distribution within a fixed range. For the IEEE 30-bus system, the scaling factors were sampled from [0.2, 1.4]; for the 200-bus system, from [0.2, 2.3]; and for the 500-bus system, from [0.2, 1.8]. These ranges were selected to ensure a proportionate number of converged AC-OPF solutions across networks of different sizes. Each instance maintained a one-to-one correspondence between the selected load buses and their scaling factors, ensuring controlled but diverse load scenarios for robust OPF modeling and learning.
This randomized perturbation approach generated a rich dataset that captures a wide range of operational variability, effectively representing uncertain and dynamic load conditions in power system operation.

\subsection{Modeling of RES Units}

To realistically capture the dynamic and uncertain behavior of renewable energy sources (RES) in the dataset, selected conventional generators were redefined as RES units in the IEEE 30-bus, synthetic 200-bus, and synthetic 500-bus systems. Specifically, in the IEEE 30-bus system, the generator at Bus 13 was modeled as an RES with a rated capacity of 30 MW. In the 200-bus system, three generator units connected to Buses 65, 147, and 155 were modeled as RES units with rated capacities of 86 MW, 97 MW, and 77 MW, respectively. Similarly, in the 500-bus system, six generators located at Buses 127, 169, 224, 353, 438, and 497 were modeled as RES units with rated capacities of 50 MW, 120 MW, 158 MW, 23 MW, 61 MW, and 19 MW, respectively.
To emulate the inherent variability of renewable generation, the active power output of each RES was randomly sampled from a uniform distribution ranging from zero to its rated capacity for each OPF instance. These RES units were modeled as non-dispatchable resources, injecting their available generation into the grid at each operating point without curtailment. Furthermore, consistent with the Federal Energy Regulatory Commission (FERC) guidelines for inverter-based resources, the reactive power capability of each RES was defined to maintain a power factor range from 0.95 leading to 0.95 lagging at the point of interconnection [22].
This modeling framework ensures that the generated dataset accurately reflects the stochastic and operationally constrained characteristics of integrating renewable energy sources in modern power systems.

\section{Result and Analysis}
The summary of model parameters for training the model is shown in Table~\ref {tab: Parameters}

\begin{table}[!t]
\centering
\caption{Summary of Dataset and Model Parameters}
\label{tab: Parameters}
\begin{tabular}{lccc}
\hline
\textbf{Parameters} & \textbf{30-Bus} & \textbf{200-Bus} & \textbf{500-Bus} \\
\hline
Samples & 7623 & 7991 & 9978 \\
Load Multiplier (max) & 1.4 & 2.3 & 1.8 \\
Load Multiplier (min) & 0.2 & 0.2 & 0.2 \\
Size of Input Features: Stage 1 & 41 & 219 & 406 \\
Size of Input Features: Stage 2 & 52 & 292 & 620 \\
Hidden Layer & 1 & 1 & 1 \\
Neurons in Hidden Layer & 90 & 400 & 400 \\
Learning Rate & 0.008 & 0.004 & 0.005 \\
Prior Scale & 0.5 & 0.5 & 0.1 \\
Activation Function & Tanh & Tanh & Tanh \\
\hline
\end{tabular}
\end{table}

\subsection{Stage 1 Results for 30 Bus}
The training and testing outcomes for the 30-bus system in Stage 1, which focuses on predicting generator set points, are summarized in Figures 3–8. Figure 3 illustrates the loss function for Stage 1, and Figure 4 presents the loss function for Stage 2, demonstrating the prediction accuracy for both active and reactive power set points, as well as voltage magnitude/voltage angle for the load flow analysis in the AC-OPF of the IEEE 30-bus system. 
Figure 5 presents a detailed comparison between the actual and predicted active power set points for both the training and testing datasets. The plots also include the ±2σ uncertainty band, which quantifies the confidence level associated with the model’s predictions. This uncertainty representation provides valuable insight into the reliability and dispersion of the predicted results under varying operating conditions.
In addition, Figure 6 displays a parity plot of the training and testing outcomes for all generator buses, visually illustrating the degree of agreement between the predicted and actual active power values. Strong alignment of data points along the diagonal line indicates high prediction accuracy and minimal bias in the model output.
The corresponding results for reactive power predictions are presented in Figures 7 and 8, which encompass both training and testing datasets. Together, these figures complement the analysis of active power predictions, offering a comprehensive view of the model's performance across different operating ranges.
All evaluations were carried out for the six generators connected to the IEEE 30-bus test system. The total number of samples, along with the count of training and testing predictions lying within the ±2σ uncertainty band, are explicitly reported in the header of each graph to ensure transparency and reproducibility. The results show that more than 95\% of both training and testing predictions fall within the ±2σ uncertainty range, underscoring the robustness and stability of the proposed model. This consistently high coverage across datasets confirms that the model achieves not only high accuracy but also strong generalization capability under diverse system operating conditions.

The ML/AI model has also been evaluated using data that include out-of-bound cases for the AC-OPF, in order to examine whether the model can adapt to conditions beyond its original training scope when predicting generation set points. In this context, out-of-bound data refer to load variations within the range [0.2, 1.6]. If the observed predictions within the ±2σ uncertainty band fall below 90\%, the model undergoes retraining within this range to improve accuracy, a process referred to as the on-the-fly approach. The results for the 30-bus system, which encompass out-of-bound testing, retraining with the on-the-fly method, and subsequent validation, are summarized in Table~\ref {tab: on-the-fly}. These results reveal that when the trained model is tested on out-of-bound data, the prediction of reactive power of the generator connected to bus 23 falls below the 90\% threshold within the ±2σ confidence band. To restore accuracy, the model generates additional AC-OPF data within the critical range (e.g., [0.2, 1.6]) and retrains on this dataset while preserving the structure and features of the original model. After retraining, the model is again tested with another dataset in this extended range, and the outcomes confirm that accuracy is successfully maintained above 95\% within the ±2σ uncertainty band.

\subsection{Stage 2 Results for 30 Bus}
Once the generator active and reactive power set points for the optimal AC power flows are established, the model proceeds to Stage 2, which involves training and testing for load flow analysis to predict bus voltages and voltage angles. Figures 9 and 10 display the trained and predicted results for nine randomly selected buses. The findings indicate that more than 90\% of both voltage magnitudes and voltage angles lie within the ±2σ confidence band. This demonstrates that, provided Stage 1 delivers sufficiently accurate generation set points, the on-the-fly retraining approach is not required in Stage 2. The robustness of Stage 2 predictions thus highlights the effectiveness of the two-stage framework in maintaining high reliability across diverse operating conditions.

\begin{table}[h!]
\centering
\caption{Training, testing, and retraining with on-the-fly approach in Stage 1 for the IEEE 30-Bus System}
\label{tab: on-the-fly}
\renewcommand{\arraystretch}{1.2}
\begin{tabular}{|l|l|c|c|c|c|c|c|}
\hline
\multicolumn{8}{|c|}{Train [0.2, 1.4]} \\ \hline
\multicolumn{2}{|l|}{Generation Bus} & 27 & 1 & 23 & 2 & 13 & 22 \\ \hline
P\_g & \multirow{2}{*}{\shortstack[c]{\% within \\ $\pm 2\sigma$ bound}} & 99.68 & 98.41 & 98.85 & 99.06 & 100 & 98.57 \\ \cline{1-1} \cline{3-8}
Q\_g & & 99.51 & 98.95 & 97.92 & 99.21 & 99.96 & 97.81 \\ \hline

\multicolumn{8}{|c|}{Test 1 [0.2, 1.4]} \\ \hline
\multicolumn{2}{|l|}{Generation Bus} & 27 & 1 & 23 & 2 & 13 & 22 \\ \hline
P\_g & \multirow{2}{*}{\shortstack[c]{\% within \\ $\pm 2\sigma$ bound}} & 98.96 & 97.05 & 97.14 & 96.81 & 100 & 96.34 \\ \cline{1-1} \cline{3-8}
Q\_g & & 99.28 & 95.06 & 96.26 & 96.97 & 100 & 97.61 \\ \hline

\multicolumn{8}{|c|}{Test 2 [0.2, 1.6]} \\ \hline
\multicolumn{2}{|l|}{Generation Bus} & 27 & 1 & 23 & 2 & 13 & 22 \\ \hline
P\_g & \multirow{2}{*}{\shortstack[c]{\% within \\ $\pm 2\sigma$ bound}} & 97.44 & 97.09 & 91.81 & 96.76 & 100 & 92.15 \\ \cline{1-1} \cline{3-8}
Q\_g & & 98.12 & 90.35 & 84.21 & 96.93 & 100 & 94.71 \\ \hline

\multicolumn{8}{|c|}{Retrain [0.2, 1.6]} \\ \hline
\multicolumn{2}{|l|}{Generation Bus} & 27 & 1 & 23 & 2 & 13 & 22 \\ \hline
P\_g & \multirow{2}{*}{\shortstack[c]{\% within \\ $\pm 2\sigma$ bound}} & 99.83 & 99.32 & 99.57 & 99.49 & 100 & 99.57 \\ \cline{1-1} \cline{3-8}
Q\_g & & 100 & 99.48 & 99.14 & 99.48 & 100 & 99.06 \\ \hline

\multicolumn{8}{|c|}{Re-test [0.2, 1.6]} \\ \hline
\multicolumn{2}{|l|}{Generation Bus} & 27 & 1 & 23 & 2 & 13 & 22 \\ \hline
P\_g & \multirow{2}{*}{\shortstack[c]{\% within \\ $\pm 2\sigma$ bound}} & 99.32 & 98.89 & 98.89 & 99.06 & 100 & 97.78 \\ \cline{1-1} \cline{3-8}
Q\_g & & 99.74 & 97.86 & 97.35 & 98.81 & 99.91 & 98.12 \\ \hline
\end{tabular}
\end{table}

\subsection{Results for Synthetic 200 Bus and 500 Bus System}
To further validate the robustness and scalability of the proposed PowerModel-ACOPF-AI framework, the model was extended and trained on larger and more complex test systems, namely the IEEE 200-bus and 500-bus networks. This evaluation aimed to examine the model’s ability to maintain high prediction accuracy and generalization capability as the system size and data dimensionality increase. 
Similar to the IEEE 30-bus case, the AC-OPF solution process was divided into two stages. Stage 1 focused on predicting the active and reactive power generation set points for all generator buses, while Stage 2 concentrated on estimating the corresponding bus voltage magnitudes and voltage phase angles. 
The detailed results for the IEEE 200-bus system are presented in Figures 11–14. Figure 11 illustrates the comparison between actual and predicted active power generation for six randomly selected generator units in both the training and testing datasets, whereas Figure 12 shows the corresponding results for reactive power generation. In total, 7,991 samples were used for training the Stage 1 and Stage 2 models, and 1,585 samples were reserved for testing Stage 1. The parity plots shows that more than 95\% of the predictions for both active and reactive power fall within the ±2σ uncertainty band, confirming the high reliability and consistency of the model’s predictions.
Once the power generation set points were obtained from Stage 1, they were used as inputs for Stage 2 to predict the bus voltage magnitudes and voltage angles, employing the same training dataset size as in Stage 1. For this stage, 1,433 samples were utilized for testing. Figure 13 depicts the actual and predicted voltage magnitudes for nine randomly selected buses, while Figure 14 presents the corresponding voltage angle predictions. The results reveal that over 95\% of the testing predictions lie within the ±2σ confidence band, further underscoring the robustness and precision of the proposed model even under the increased complexity of the 200-bus system.

The framework was then further validated on an even larger system, the IEEE 500-bus test case, to assess its scalability under more complex operational scenarios. To represent variability from renewable integration, six of the 60 generating units in this system were modeled as renewable sources, with output power fluctuating randomly within their respective operating limits. As with the smaller cases, the AC-OPF was performed in two stages: Stage 1 for predicting active and reactive power set points, and Stage 2 for computing the resulting bus voltages and phase angles. For this system, 9,978 samples were used for training and 1,996 samples were used for testing across both stages. The Stage 1 results for six randomly selected generators are presented in Figures 15 and 16, showing that more than 95\% of both active and reactive power predictions fall within the ±2σ uncertainty band, even under renewable intermittency. In Stage 2, the predicted and actual results for nine randomly chosen buses are shown in Figures 17 and 18. These results confirm that the proposed two-stage model consistently maintains high prediction accuracy and robust uncertainty calibration, even for large-scale systems with significant renewable penetration. The strong performance across the 30-, 200-, and 500-bus networks demonstrates the scalability and adaptability of the framework, making it well-suited for practical power system applications. 

Note that the On-the-fly application has not been implemented in the results of the synthetic 200-bus and 500-bus systems. The on-the-fly learning mechanism operates in an event-triggered manner and is activated only when prediction performance degrades beyond a predefined threshold. As a result, retraining is infrequent and localized, and its overall computational burden remains limited. Moreover, the use of targeted data generation ensures that only a limited number of additional AC-OPF samples are required, avoiding the need for full retraining over the entire dataset. For larger systems, such as the 200-bus and 500-bus networks, retraining needs to perform only when necessary and does not constitute a continuous computational burden, thereby ensuring that real-time decision-making is not hindered and preserving the practical applicability of the proposed approach in time-critical power system operations.

To evaluate the robustness of the proposed framework under high variability, a wide range of operating conditions was incorporated into the dataset through stochastic perturbations of both load demand and renewable generation. Load levels were varied using scaling factors across broad intervals, while renewable outputs were randomly sampled within their rated capacities, thereby emulating realistic uncertainty in power system operation. In addition, out-of-distribution scenarios were explicitly tested by extending the load variation beyond the training range, which provided insight into the model’s sensitivity to unseen operating conditions. The results indicate that the proposed approach maintains high prediction accuracy and reliable uncertainty quantification even under significant variability. Furthermore, the on-the-fly learning mechanism enables adaptive refinement of the model in regions with degraded performance, effectively enhancing robustness against extreme or previously unobserved system states. This experimental framework serves as an implicit sensitivity analysis, demonstrating the model’s capability to generalize across a wide spectrum of uncertain operating conditions.

The proposed framework is not intended to replace conventional AC optimal power flow (AC-OPF) solvers; rather, it positions Bayesian Neural Network (BNN)-based surrogates as computational accelerators in regimes where physics-based OPF becomes increasingly challenging, such as large-scale interconnected systems. Within this context, BNN predictions serve as high-quality warm-start initializations that can significantly reduce solver runtime and improve numerical robustness. Given the strong dependence of nonlinear OPF solvers on initialization quality, the learned surrogate consistently reduces the number of required iterations, leading to meaningful computational savings, particularly for large-scale networks. Furthermore, these informed initializations enhance convergence reliability under stressed and highly uncertain operating conditions.

\begin{figure}[!t]
  \centering
  \includegraphics[width=1\linewidth]{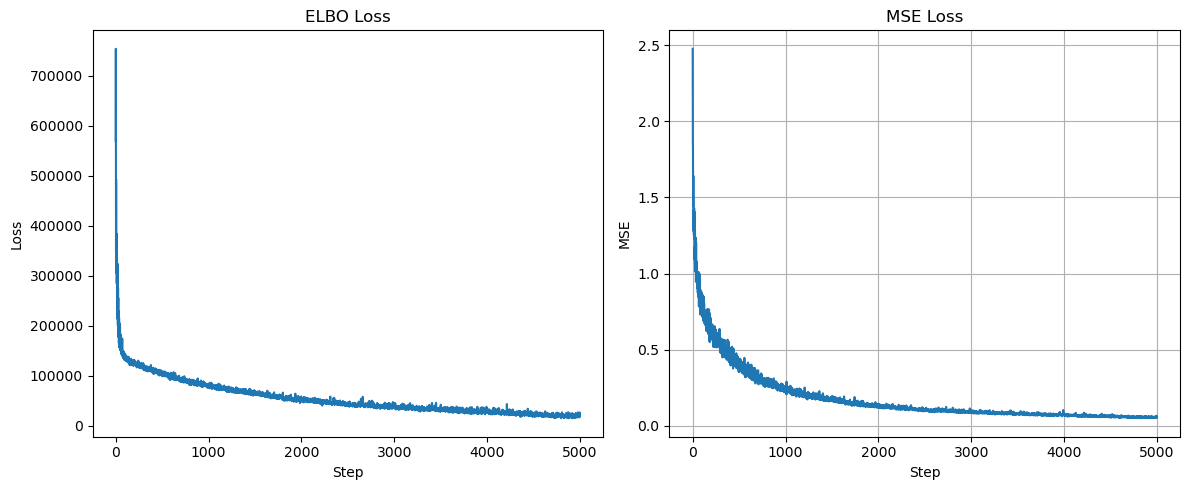}
  \caption{\textbf{Loss function for the stage 1: generation active and reactive power set point prediction for AC OPF for 30 bus system.}}
  \label{fig: Stage1 Loss}
\end{figure}

\begin{figure}[!t]
  \centering
  \includegraphics[width=1\linewidth]{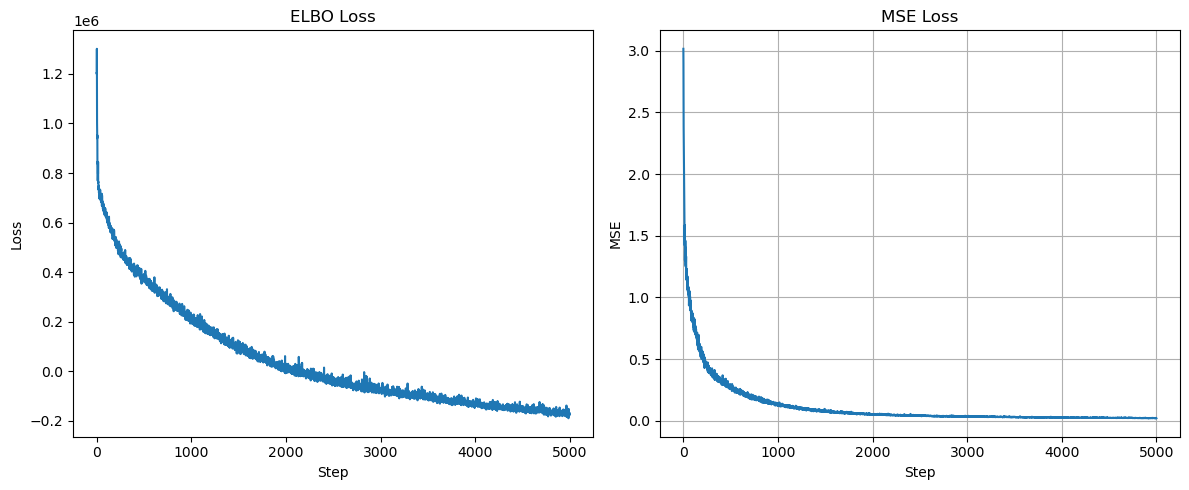}
  \caption{\textbf{Loss function for the stage 2: Voltage magnitude and angle prediction after AC OPF for 30 bus system.}}
  \label{fig: Stage1 Loss}
\end{figure}

\begin{figure*}[!t]
    \centering
    \includegraphics[width=0.48\textwidth]{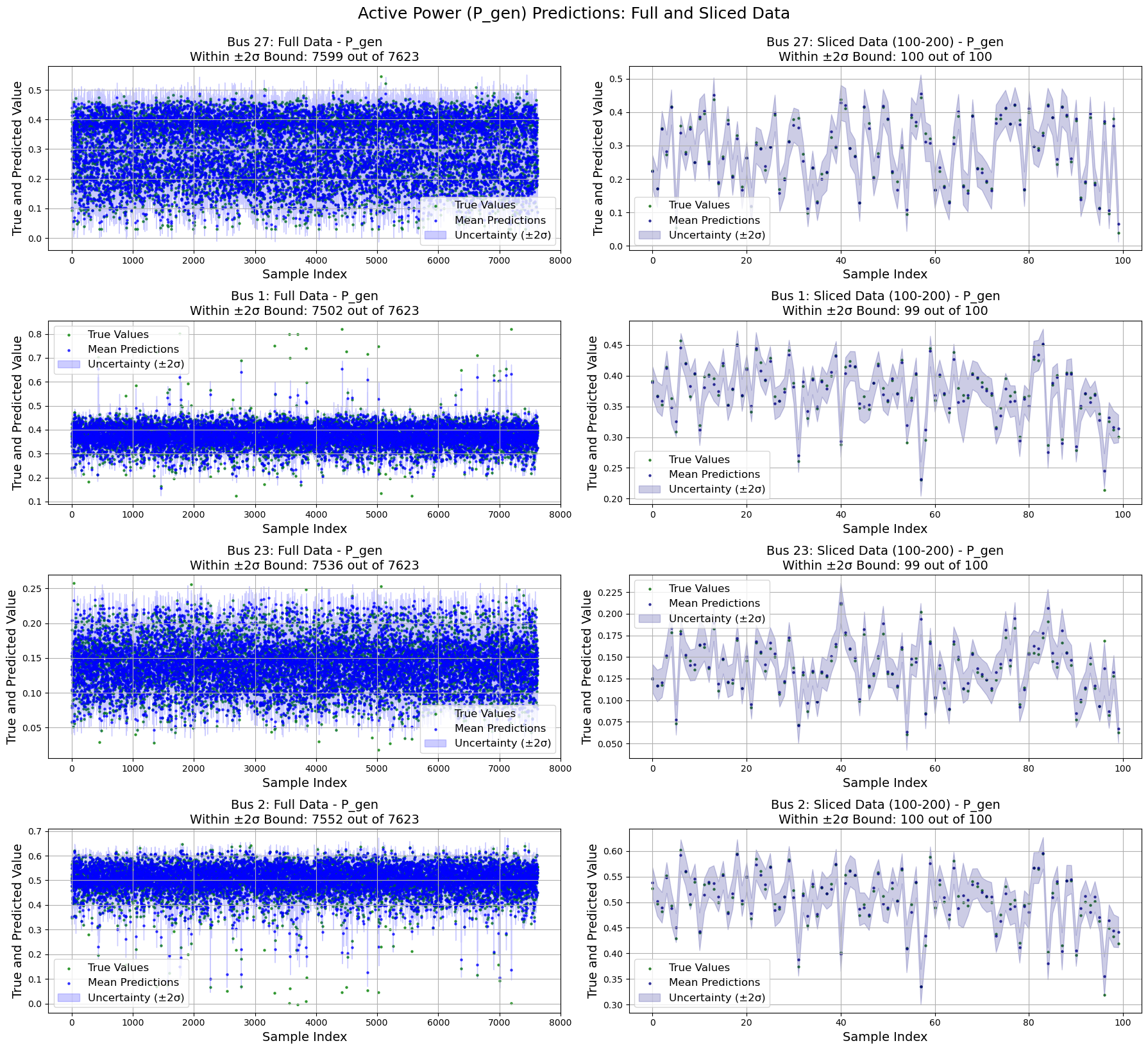}
    \hspace{1mm}
    \includegraphics[width=0.48\textwidth]{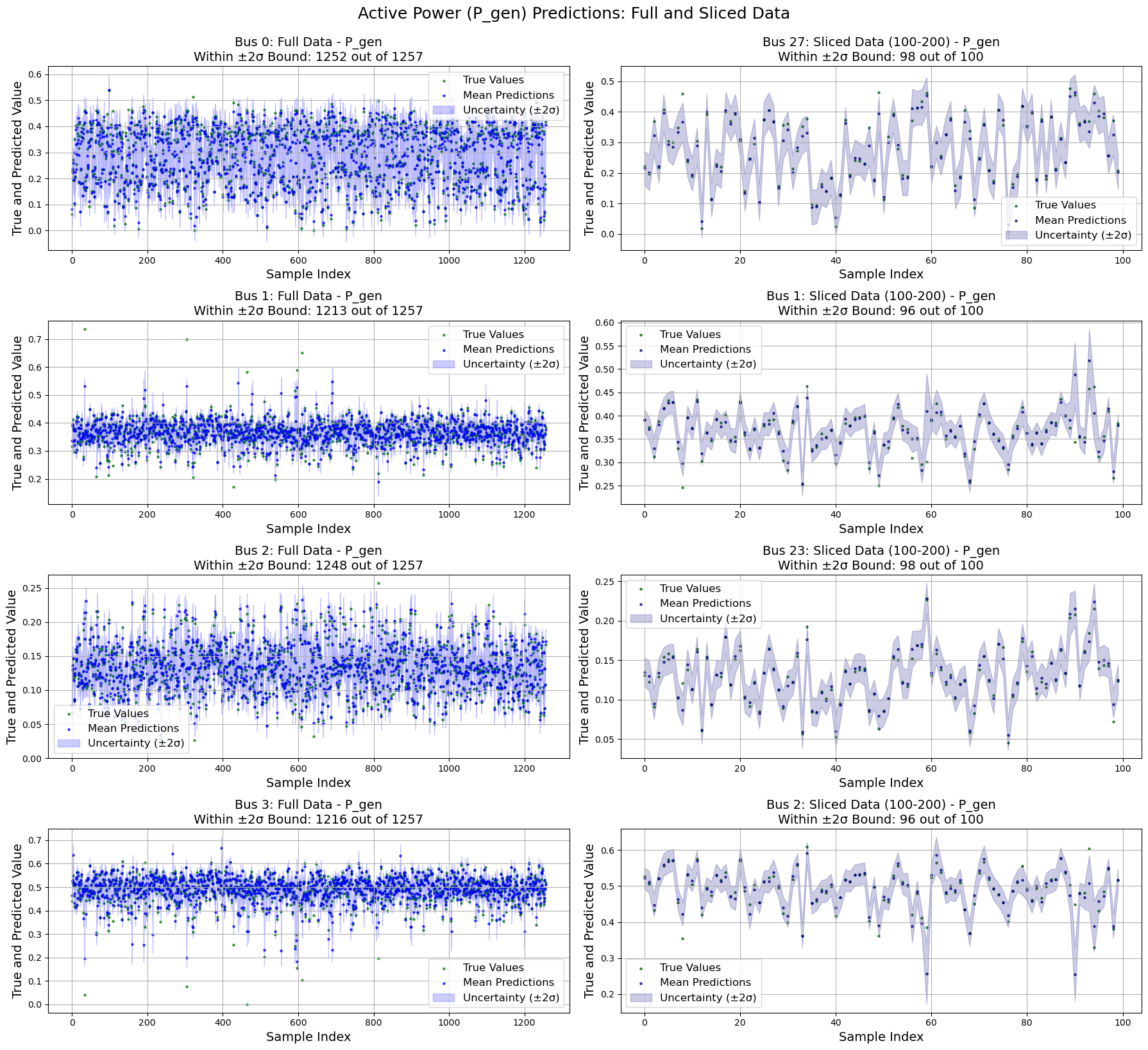}

    \par\vspace{1mm}
    {\scriptsize
    \makebox[0.48\textwidth]{\centering (a) Train}
    \hspace{1mm}
    \makebox[0.48\textwidth]{\centering (b) Test}
    }

    \caption{Actual and predicted results for train (left) and test (right) data of active power set point prediction with ±2σ uncertainty band for AC-OPF in 30 Bus system.}
    \label{fig:train_test}
    
\end{figure*}

\begin{figure*}[!t]
    \centering
    \includegraphics[width=0.48\textwidth]{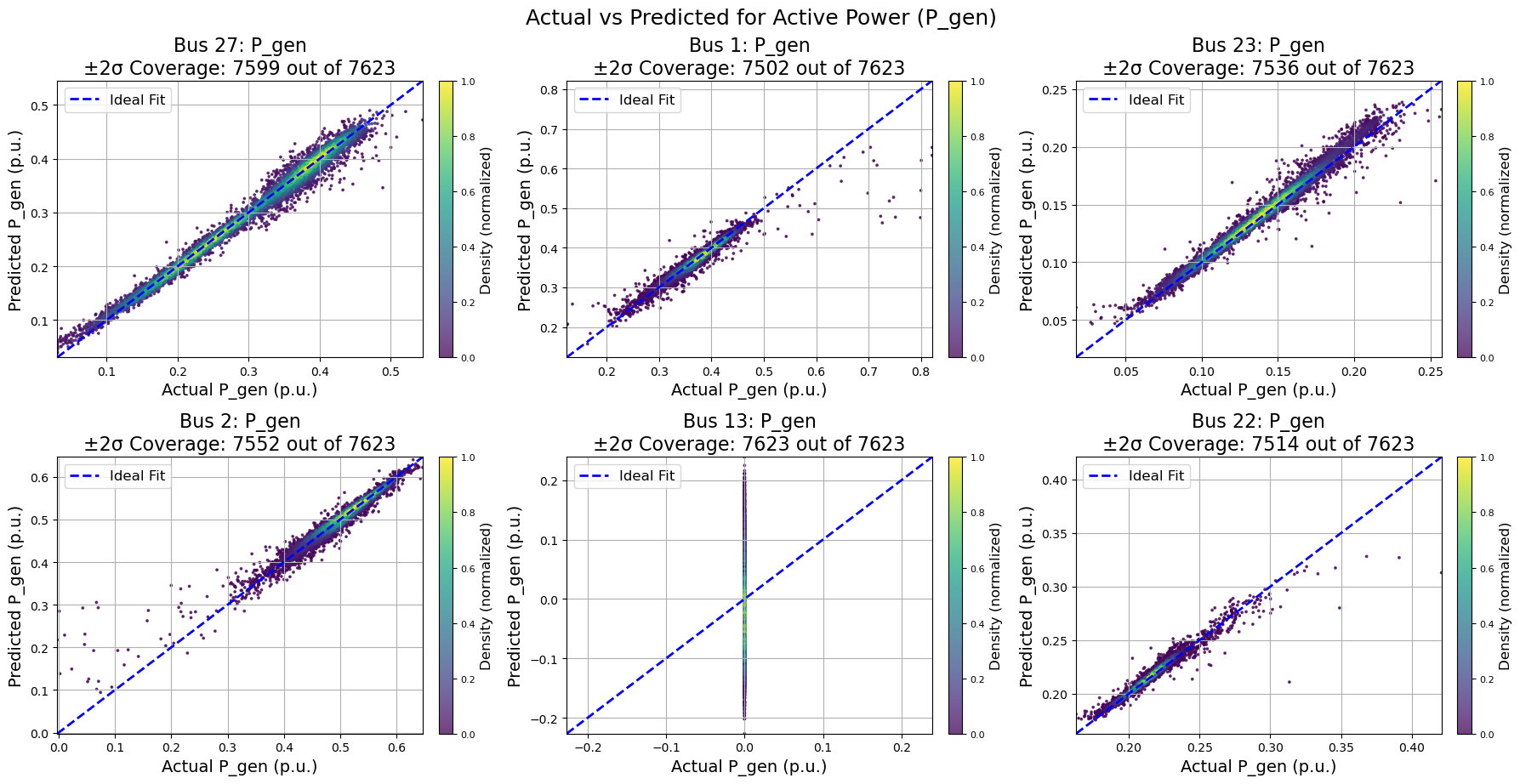}
    \hspace{1mm}
    \includegraphics[width=0.48\textwidth]{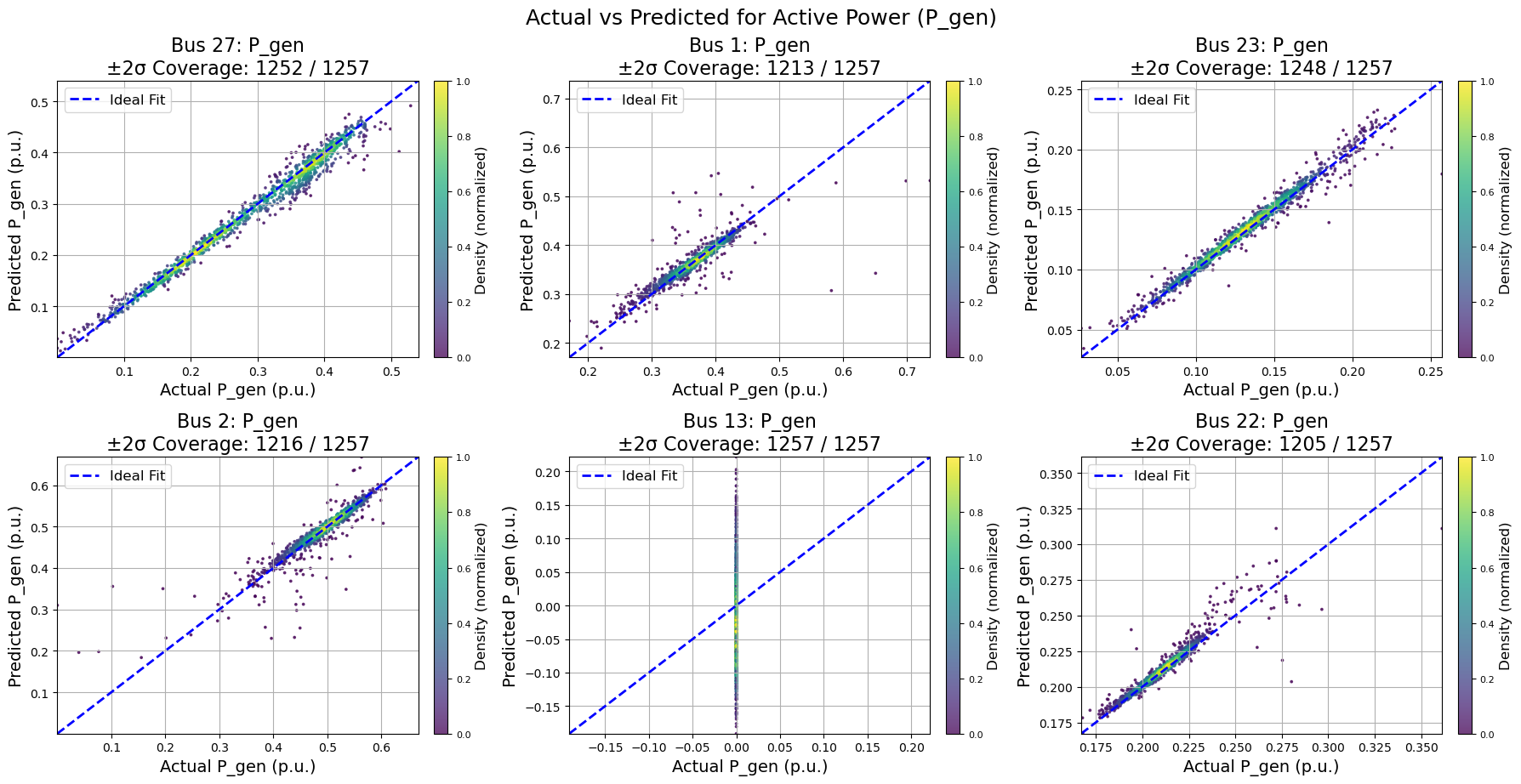}

    \par\vspace{1mm}
    {\scriptsize
    \makebox[0.48\textwidth]{\centering (a) Train}
    \hspace{1mm}
    \makebox[0.48\textwidth]{\centering (b) Test}
    }

    \caption{Parity plots from Stage 1 comparing actual and predicted active power set points for the selected buses in the synthetic 30-bus system for AC-OPF, shown for the training data (left) and testing data (right)}
    \label{fig:train_test}
\end{figure*}

\begin{figure*}[!t]
    \centering
    \includegraphics[width=0.48\textwidth]{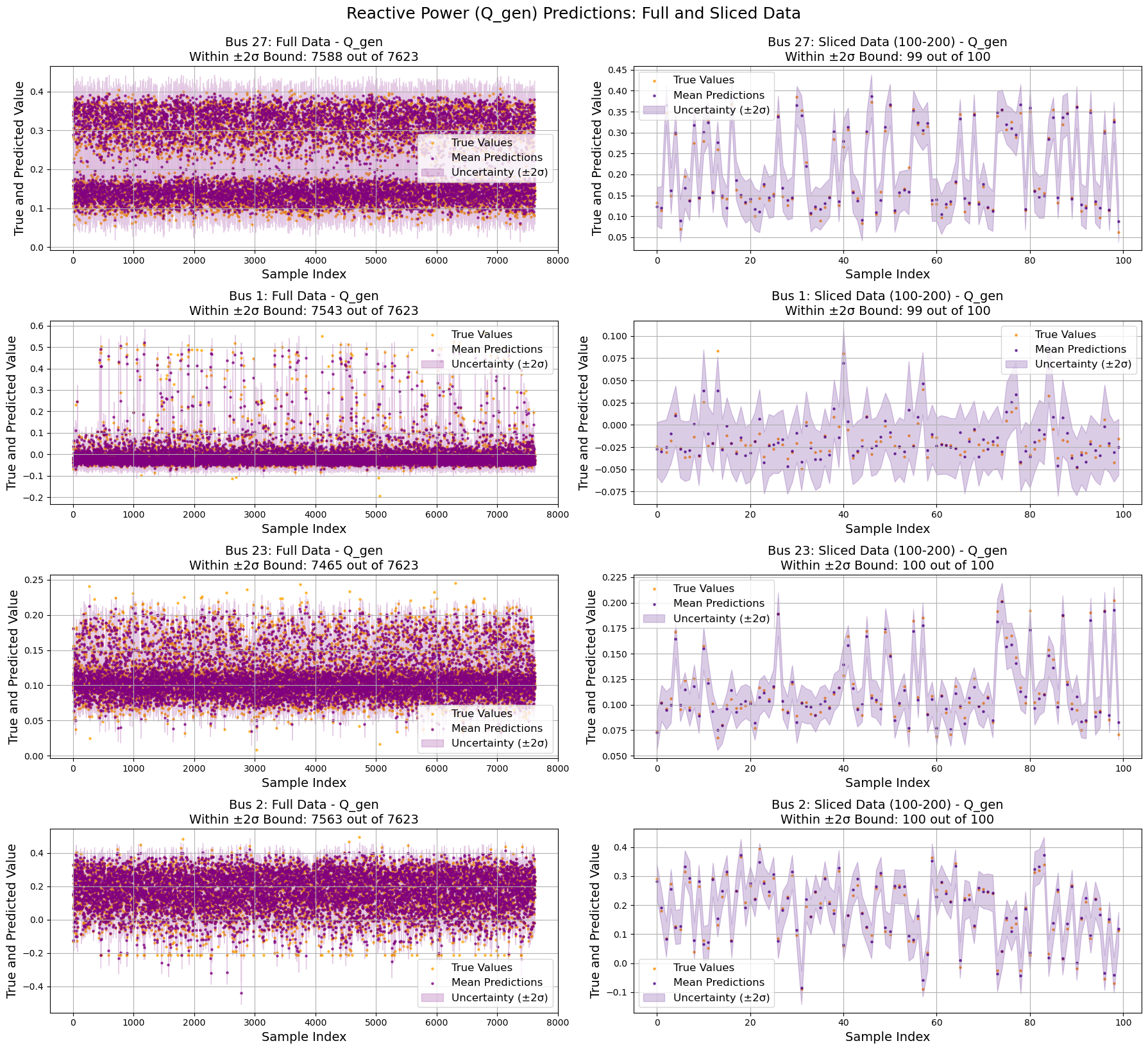}
    \hspace{1mm}
    \includegraphics[width=0.48\textwidth]{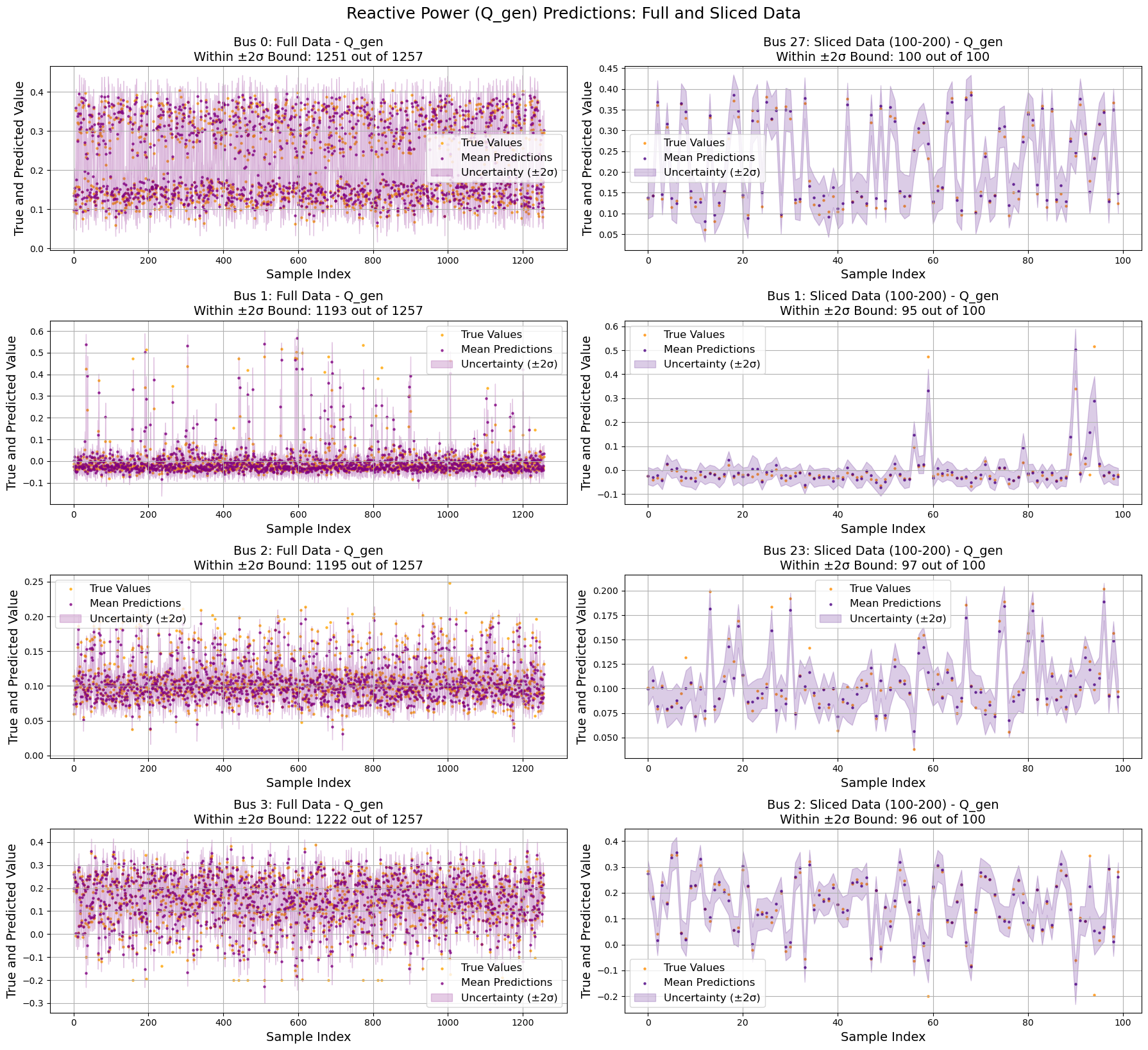}

    \par\vspace{1mm}
    {\scriptsize
    \makebox[0.48\textwidth]{\centering (a) Train}
    \hspace{1mm}
    \makebox[0.48\textwidth]{\centering (b) Test}
    }

    \caption{Actual and predicted results for train (left) and test (right) data of reactive power set point prediction with ±2σ uncertainty band for AC-OPF in 30 Bus system.}
    \label{fig:train_test}
\end{figure*}

\begin{figure*}[!t]
    \centering
    \includegraphics[width=0.48\textwidth]{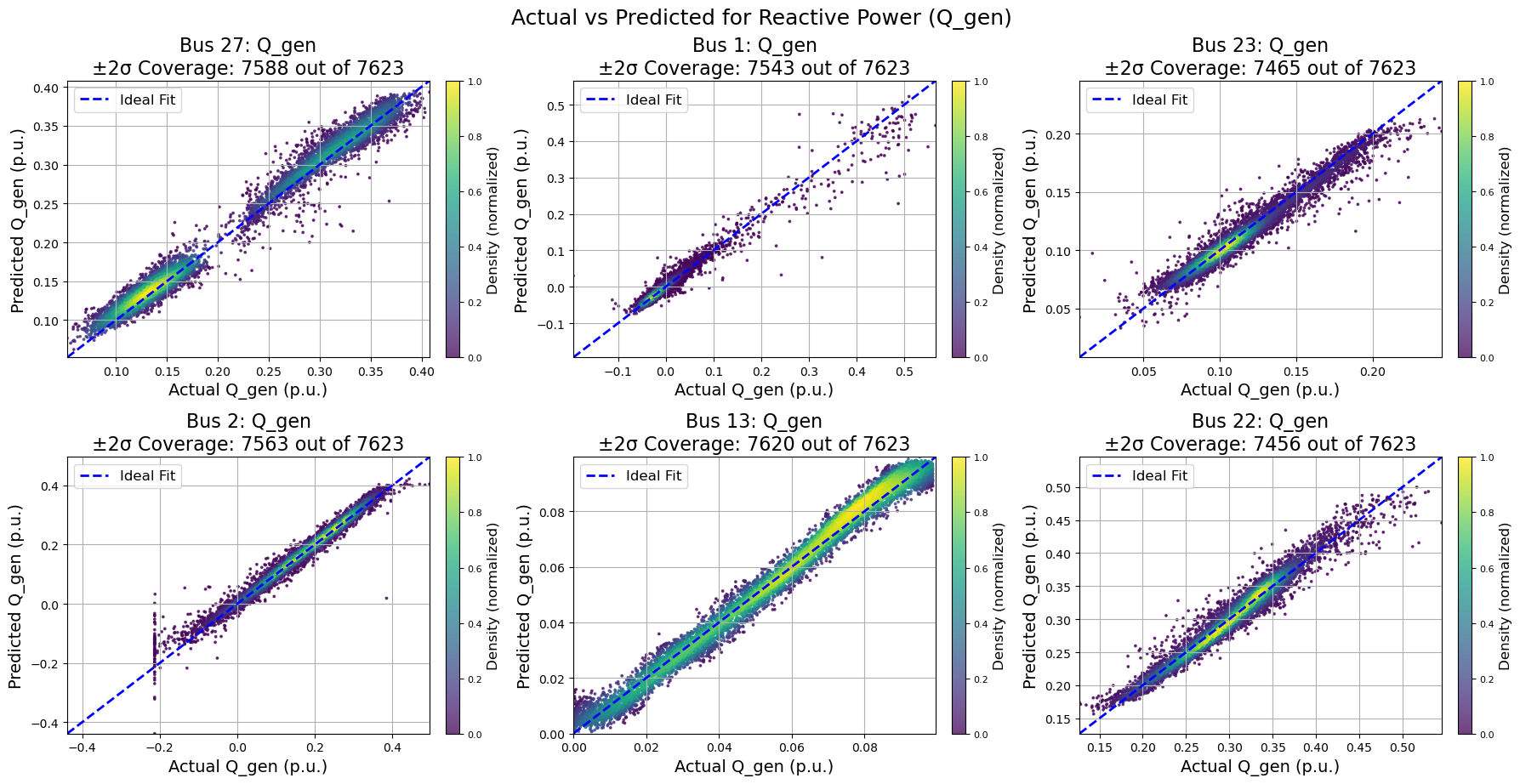}
    \hspace{1mm}
    \includegraphics[width=0.48\textwidth]{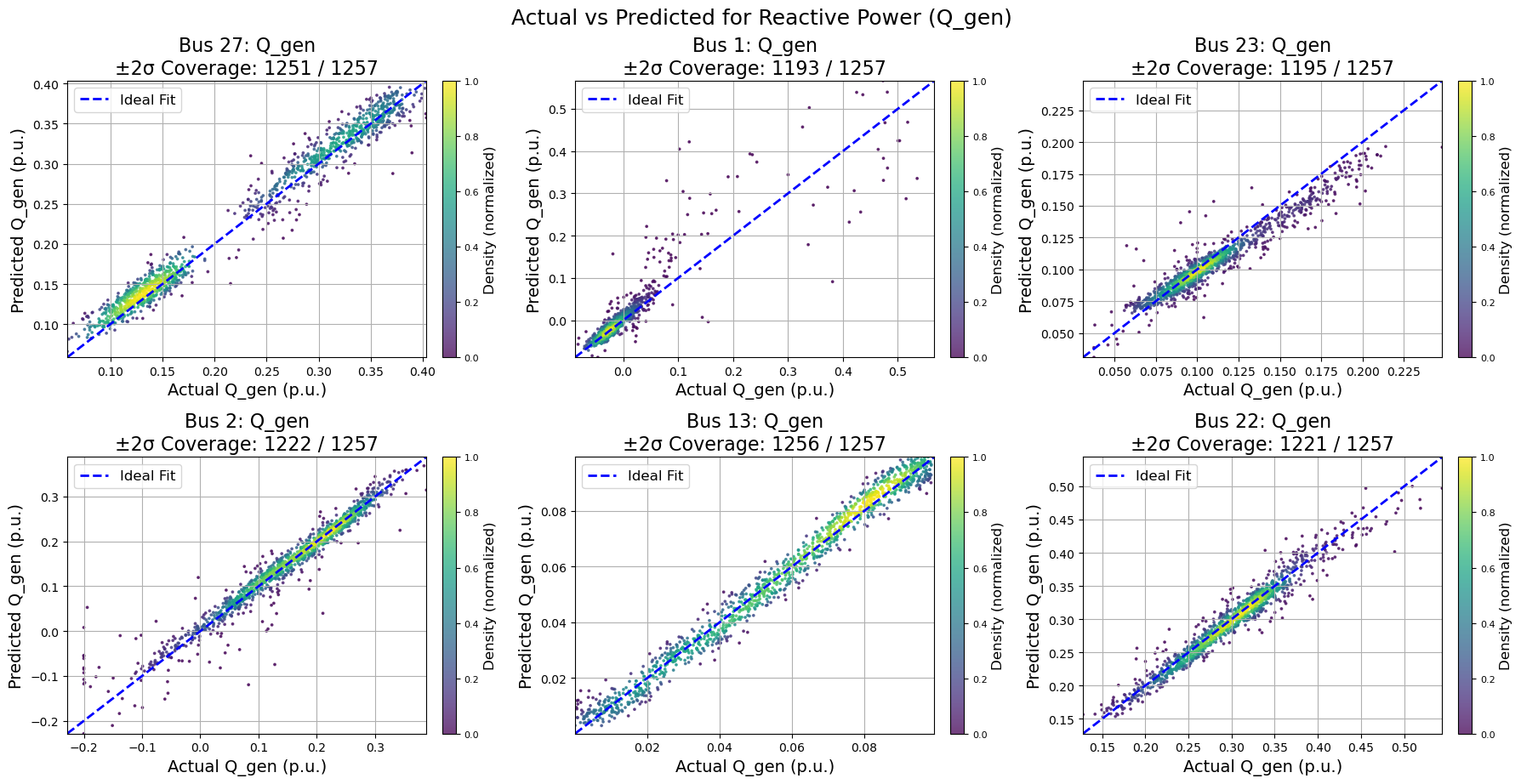}

    \par\vspace{1mm}
    {\scriptsize
    \makebox[0.48\textwidth]{\centering (a) Train}
    \hspace{1mm}
    \makebox[0.48\textwidth]{\centering (b) Test}
    }

    \caption{Parity plots from Stage 1 comparing actual and predicted reactive power set points for the selected buses in the synthetic 30-bus system for AC-OPF, shown for the training data (left) and testing data (right)}
    \label{fig:train_test}
\end{figure*}

\begin{figure*}[!t]
    \centering
    \includegraphics[width=0.48\textwidth]{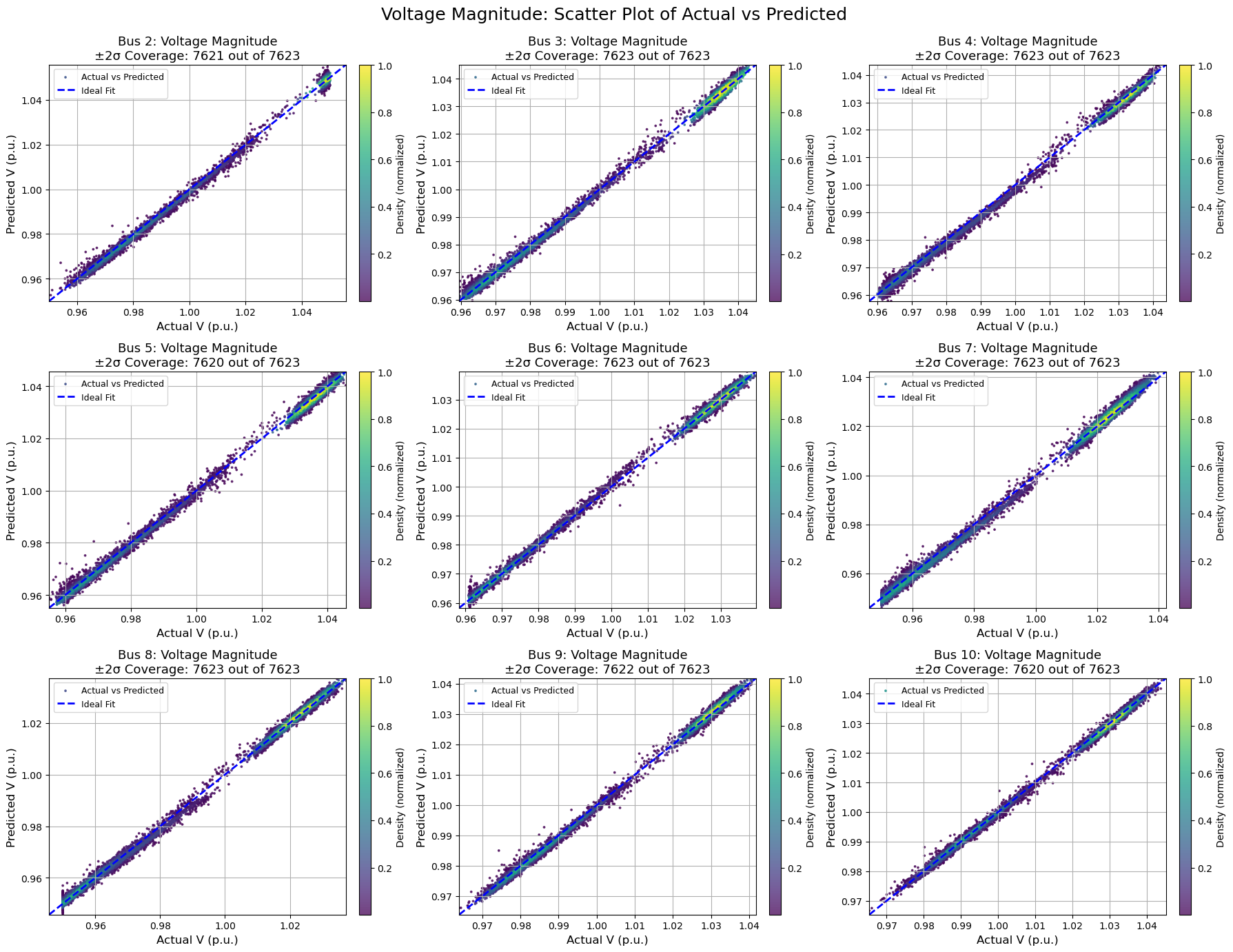}
    \hspace{1mm}
    \includegraphics[width=0.48\textwidth]{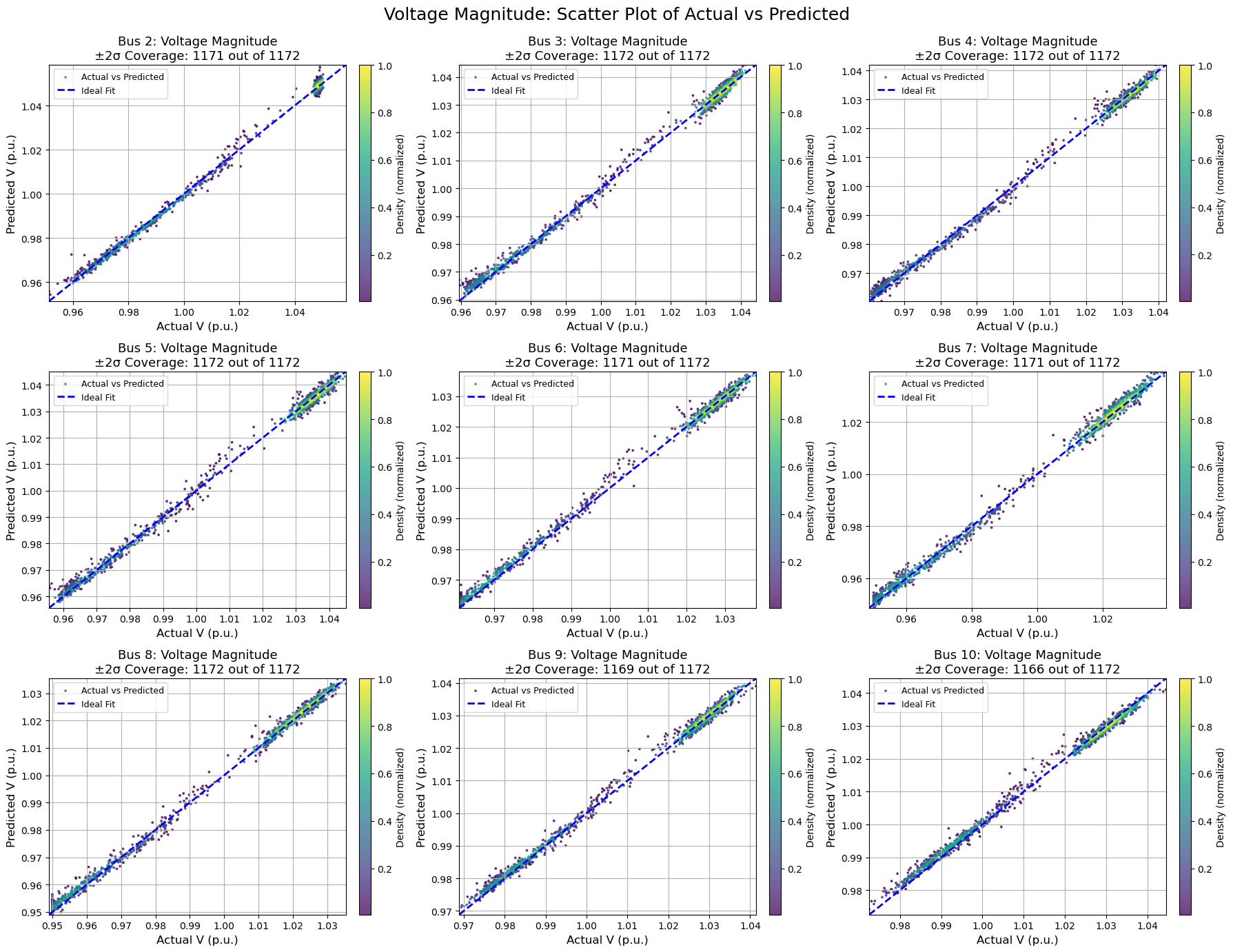}

    \par\vspace{1mm}
    {\scriptsize
    \makebox[0.48\textwidth]{\centering (a) Train}
    \hspace{1mm}
    \makebox[0.48\textwidth]{\centering (b) Test}
    }

    \caption{Parity plots from Stage 2 comparing actual and predicted results of voltage magnitude for the selected buses in the synthetic 30-bus system with AC-OPF, shown for the training data (left) and testing data (right)}
    \label{fig:train_test}
\end{figure*}

\begin{figure*}[!t]
    \centering
    \includegraphics[width=0.48\textwidth]{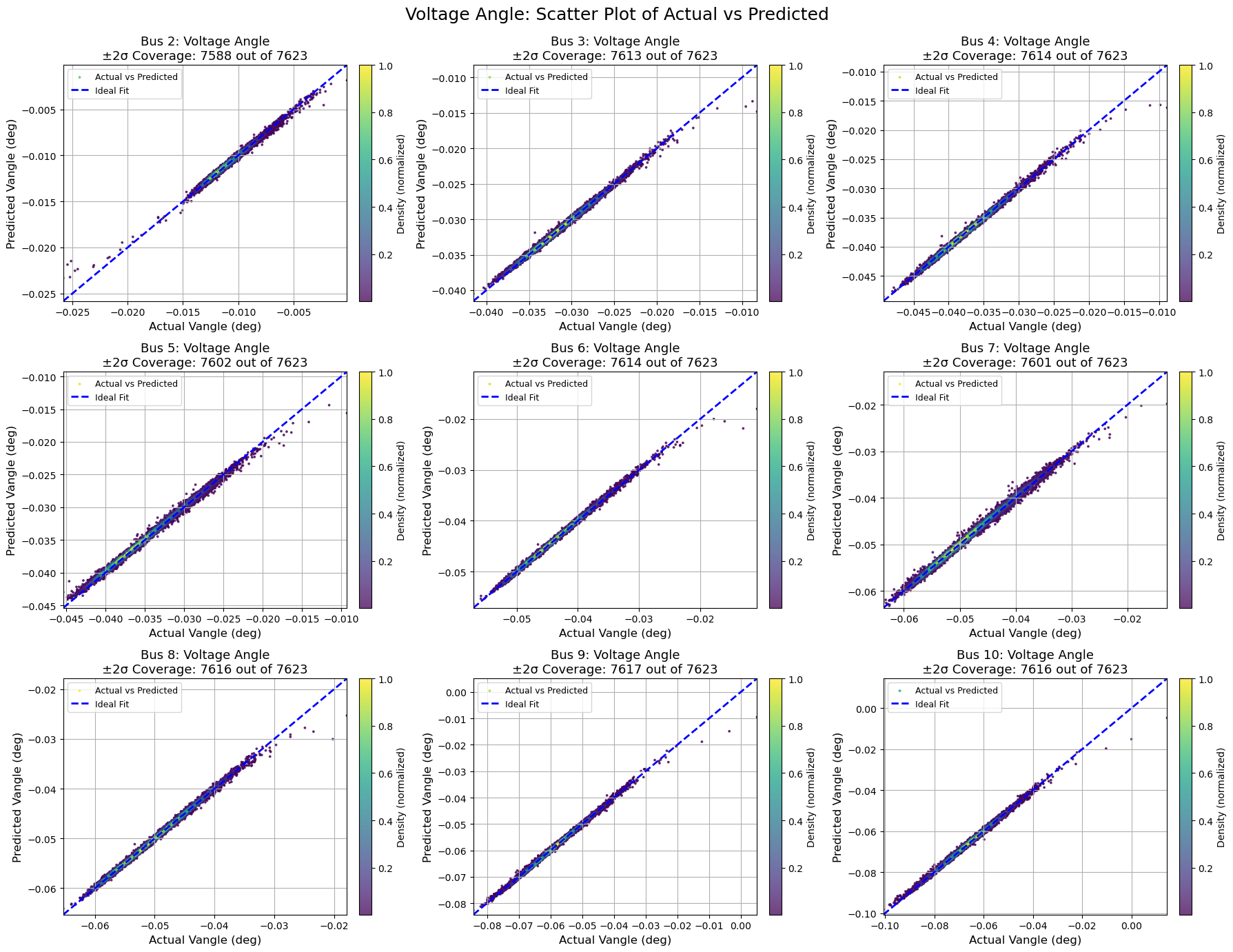}
    \hspace{1mm}
    \includegraphics[width=0.48\textwidth]{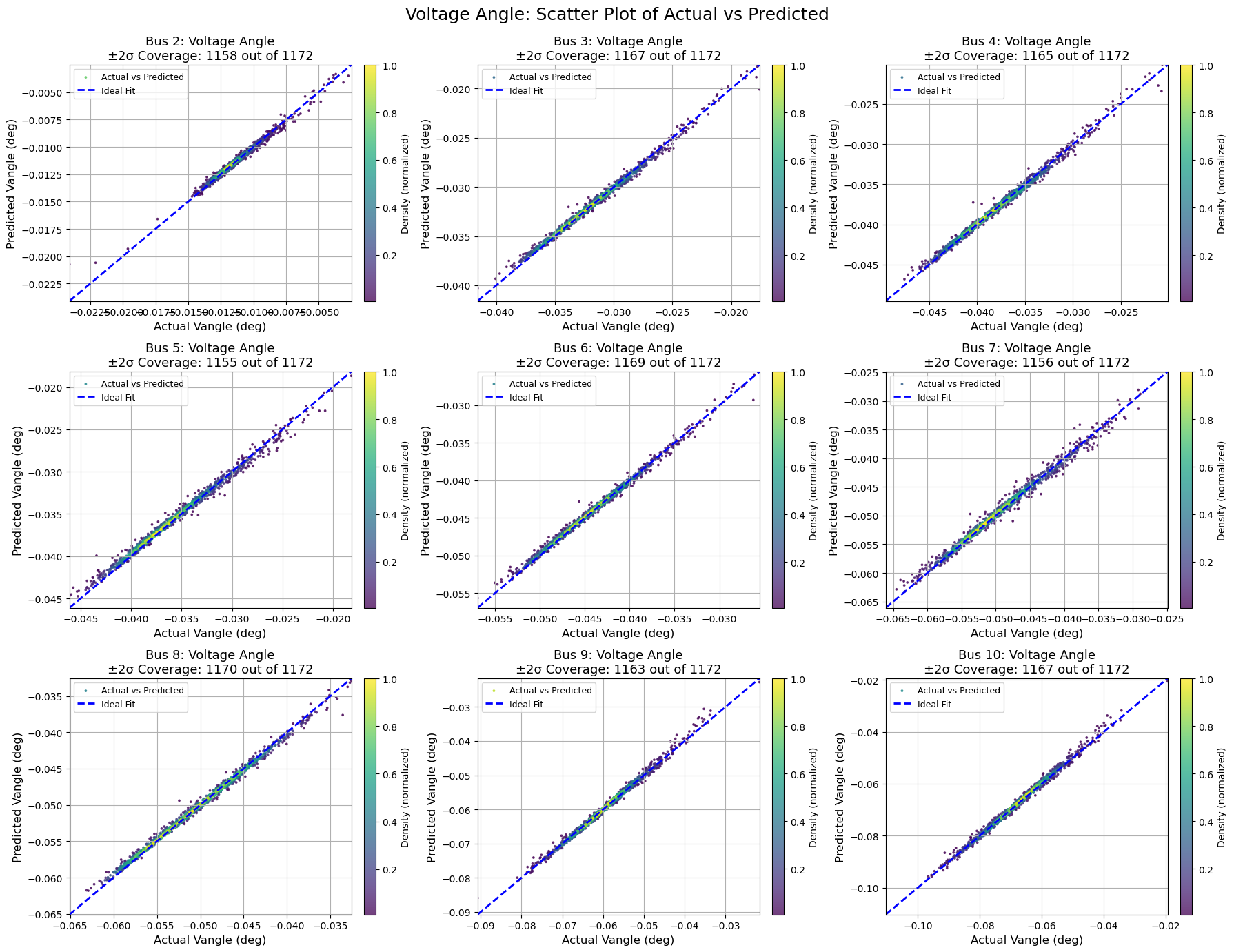}

    \par\vspace{1mm}
    {\scriptsize
    \makebox[0.48\textwidth]{\centering (a) Train}
    \hspace{1mm}
    \makebox[0.48\textwidth]{\centering (b) Test}
    }

    \caption{Parity plots from Stage 2 comparing actual and predicted results of voltage angle for the selected buses in the synthetic 30-bus system with AC-OPF, shown for the training data (left) and testing data (right)}
    \label{fig:train_test}
\end{figure*}

\begin{figure*}[!t]
    \centering
    \includegraphics[width=0.48\textwidth]{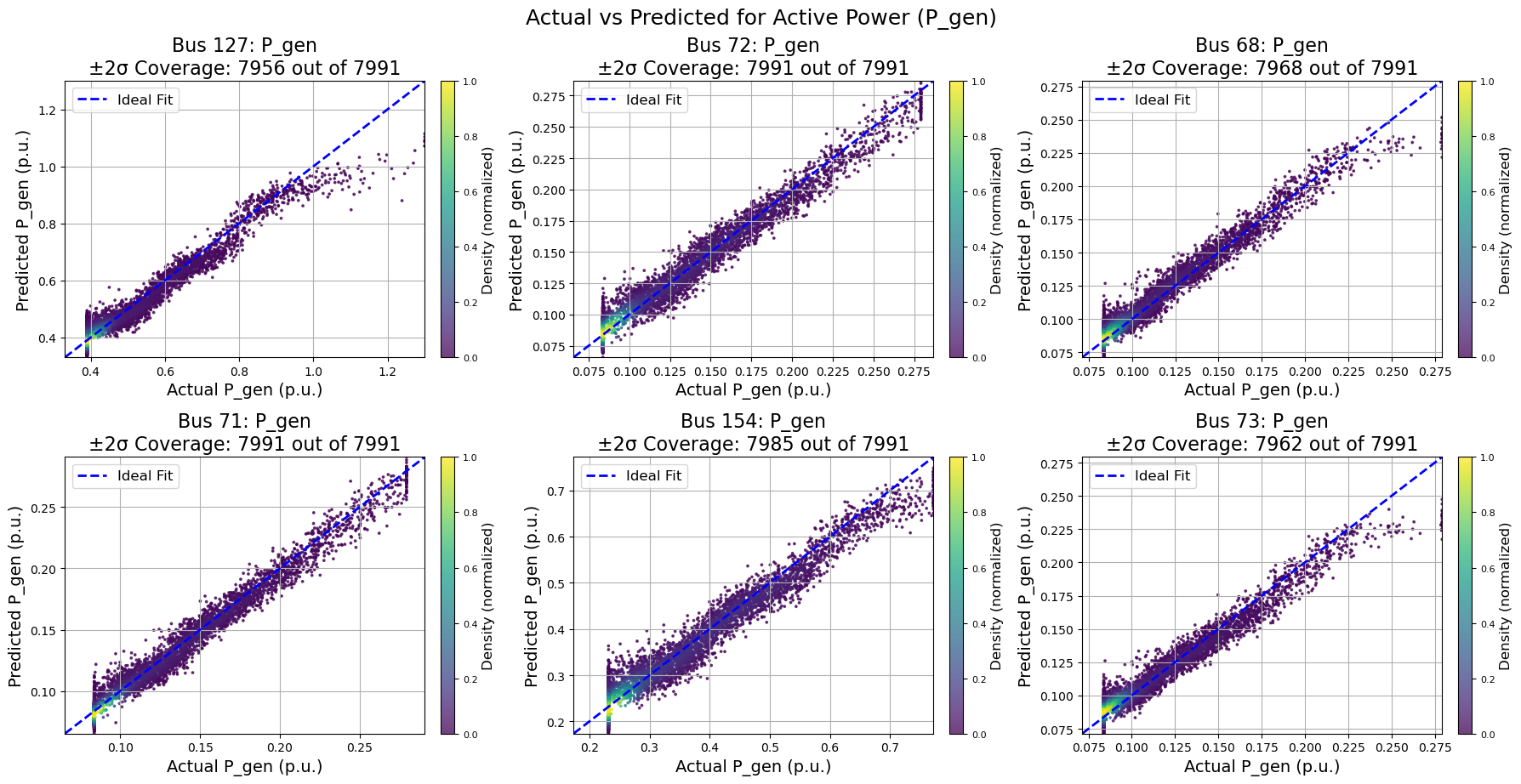}
    \hspace{1mm}
    \includegraphics[width=0.48\textwidth]{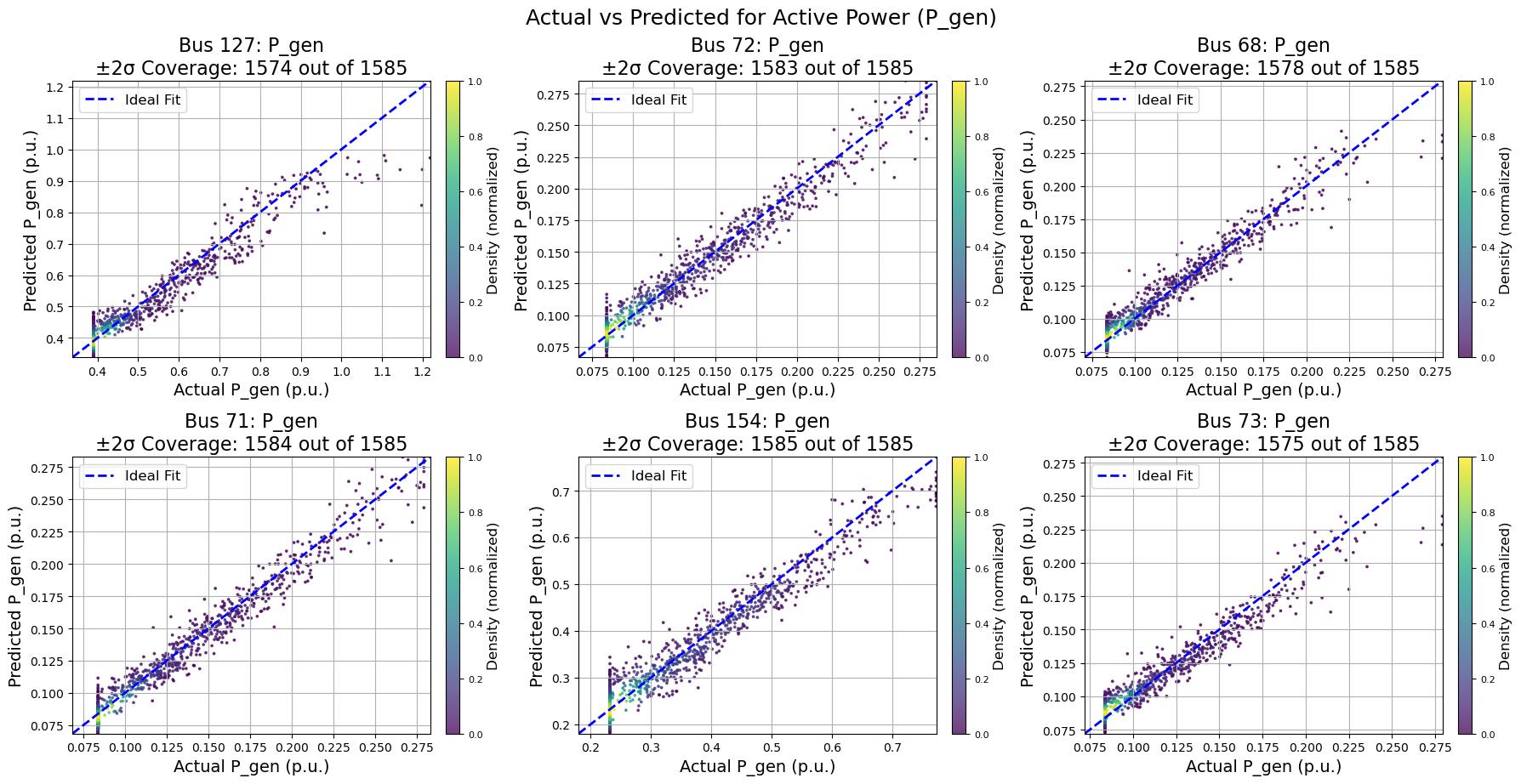}

    \par\vspace{1mm}
    {\scriptsize
    \makebox[0.48\textwidth]{\centering (a) Train}
    \hspace{1mm}
    \makebox[0.48\textwidth]{\centering (b) Test}
    }

    \caption{Parity plots from Stage 1 comparing actual and predicted active power set points for the selected buses in the synthetic 200-bus system for AC-OPF, shown for the training data (left) and testing data (right)}
    \label{fig:train_test}
\end{figure*}

\begin{figure*}[!t]
    \centering
    \includegraphics[width=0.48\textwidth]{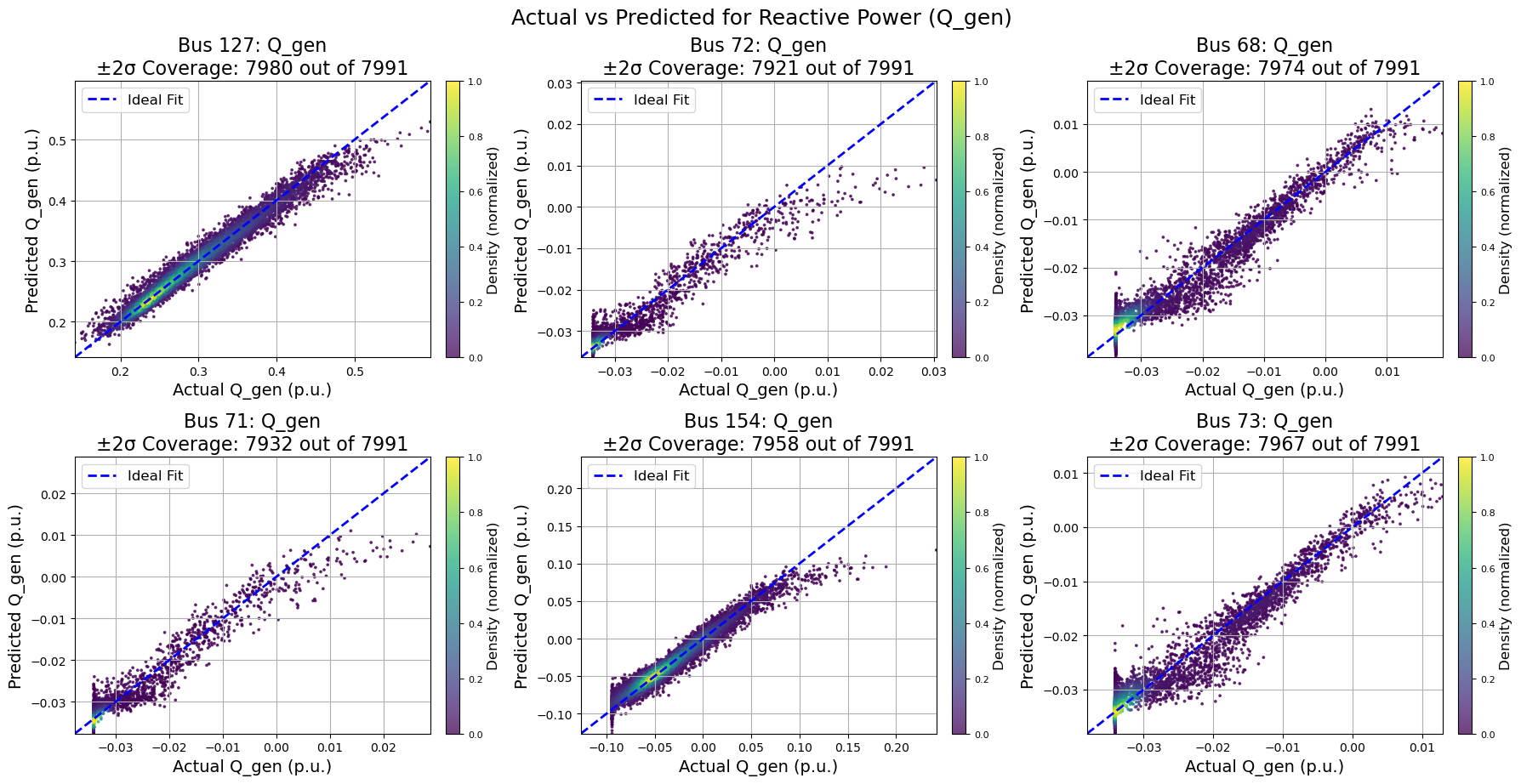}
    \hspace{1mm}
    \includegraphics[width=0.48\textwidth]{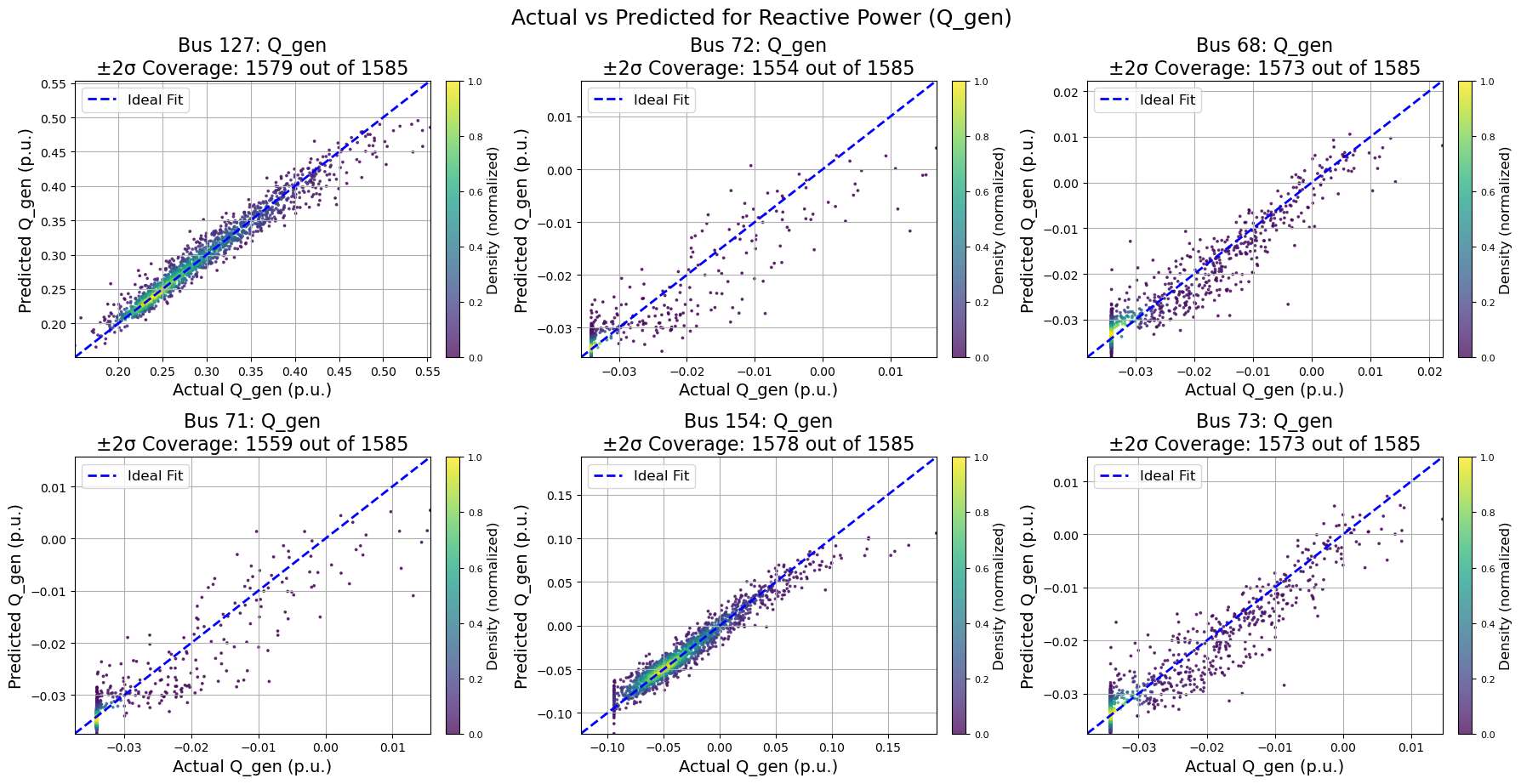}

    \par\vspace{1mm}
    {\scriptsize
    \makebox[0.48\textwidth]{\centering (a) Train}
    \hspace{1mm}
    \makebox[0.48\textwidth]{\centering (b) Test}
    }

    \caption{Parity plots from Stage 1 comparing actual and predicted reactive power set points for the selected buses in the synthetic 200-bus system for AC-OPF, shown for the training data (left) and testing data (right)}
    \label{fig:train_test}
\end{figure*}

\begin{figure*}[!t]
    \centering
    \includegraphics[width=0.48\textwidth]{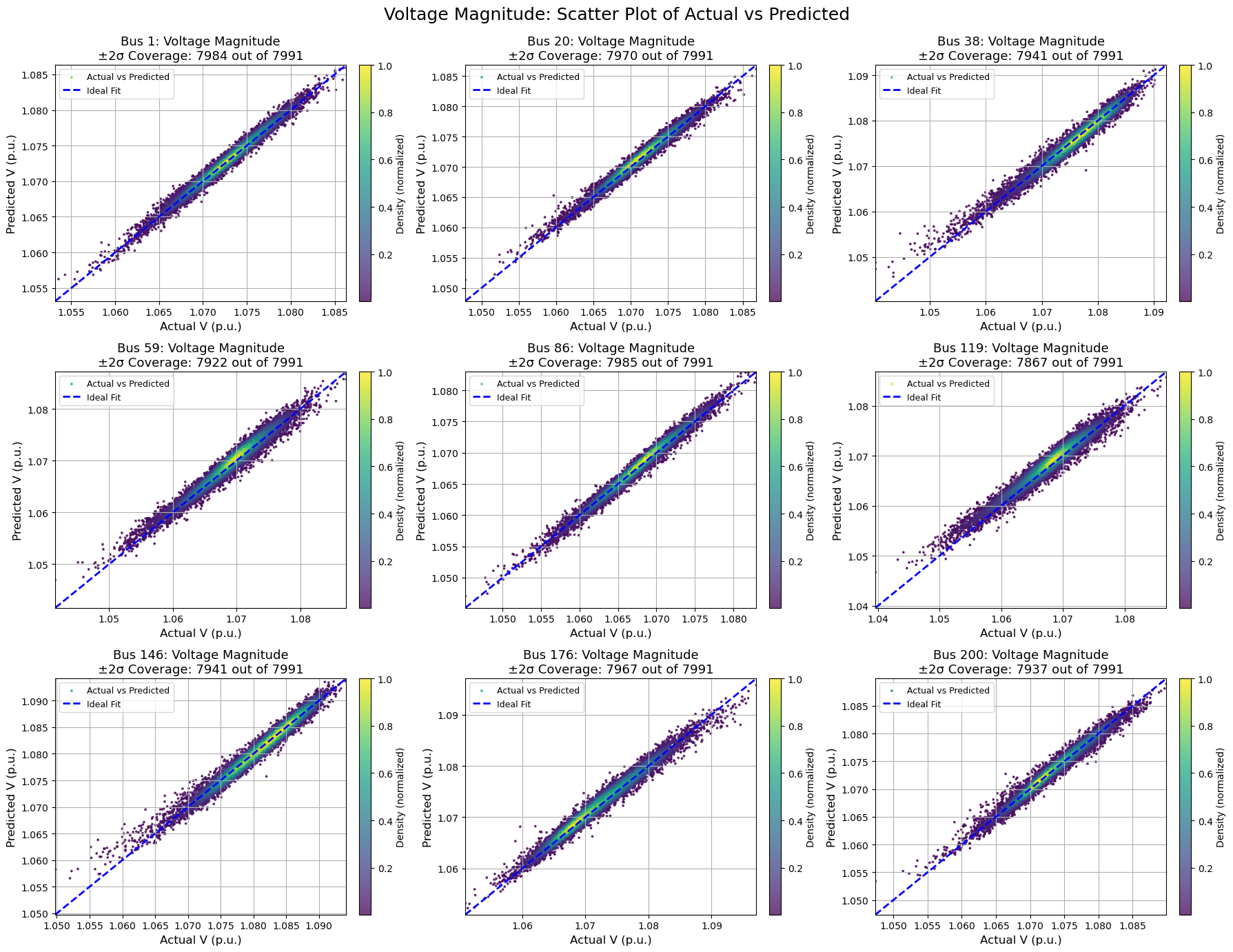}
    \hspace{1mm}
    \includegraphics[width=0.48\textwidth]{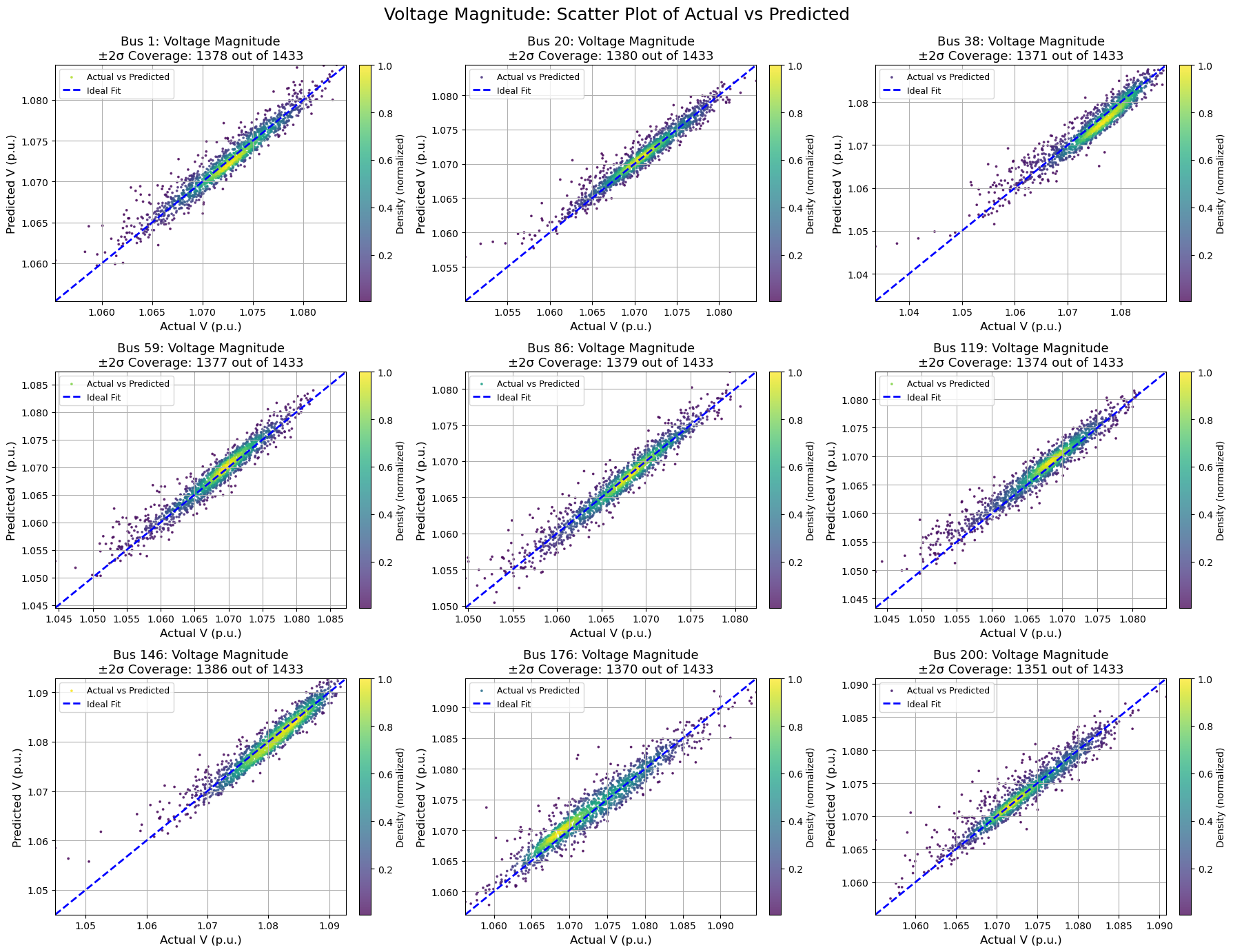}

    \par\vspace{1mm}
    {\scriptsize
    \makebox[0.48\textwidth]{\centering (a) Train}
    \hspace{1mm}
    \makebox[0.48\textwidth]{\centering (b) Test}
    }

    \caption{Parity plots from Stage 2 comparing actual and predicted results of voltage magnitude for the selected buses in the synthetic 200-bus system with AC-OPF, shown for the training data (left) and testing data (right)}
    \label{fig:train_test}
\end{figure*}

\begin{figure*}[!t]
    \centering
    \includegraphics[width=0.48\textwidth]{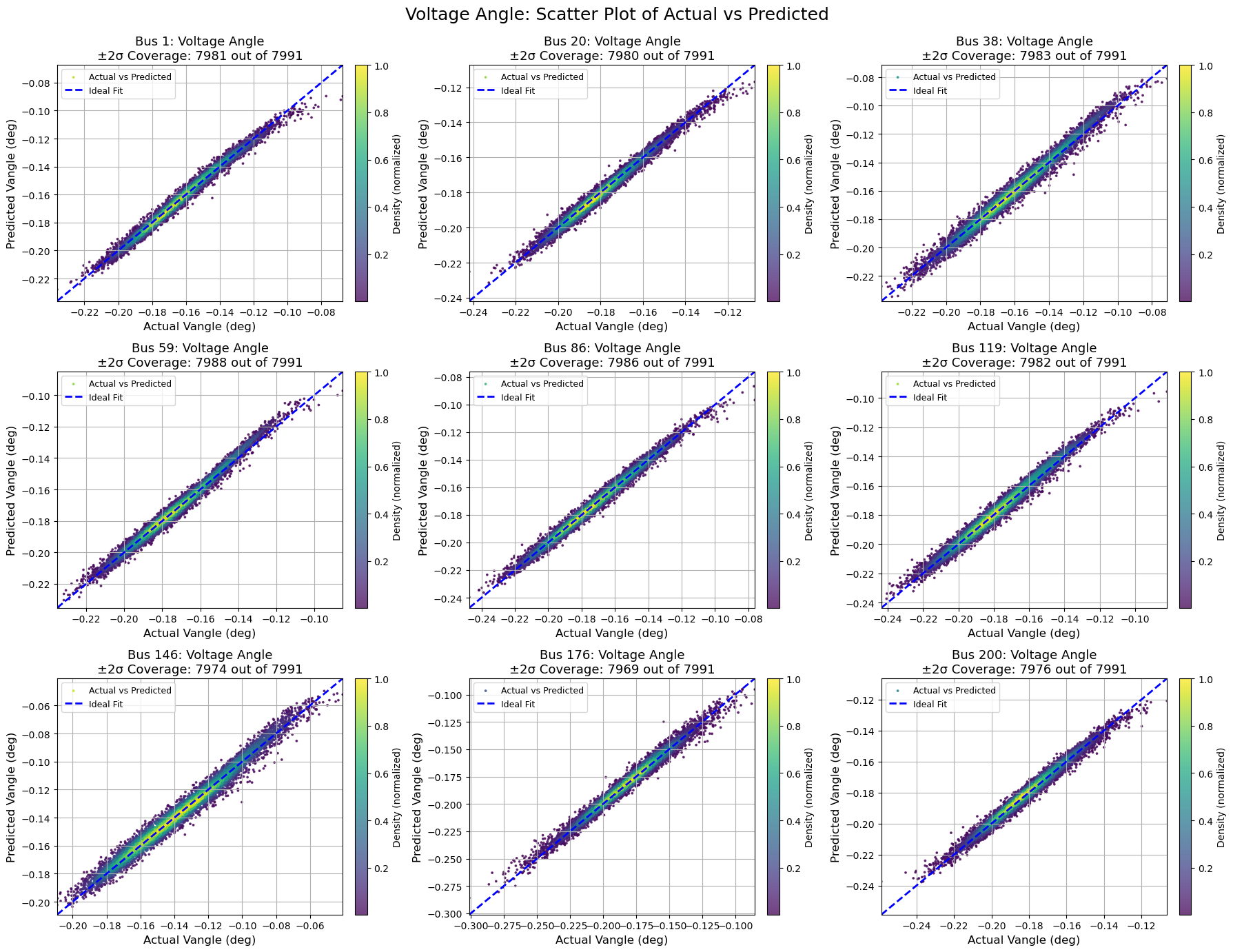}
    \hspace{1mm}
    \includegraphics[width=0.48\textwidth]{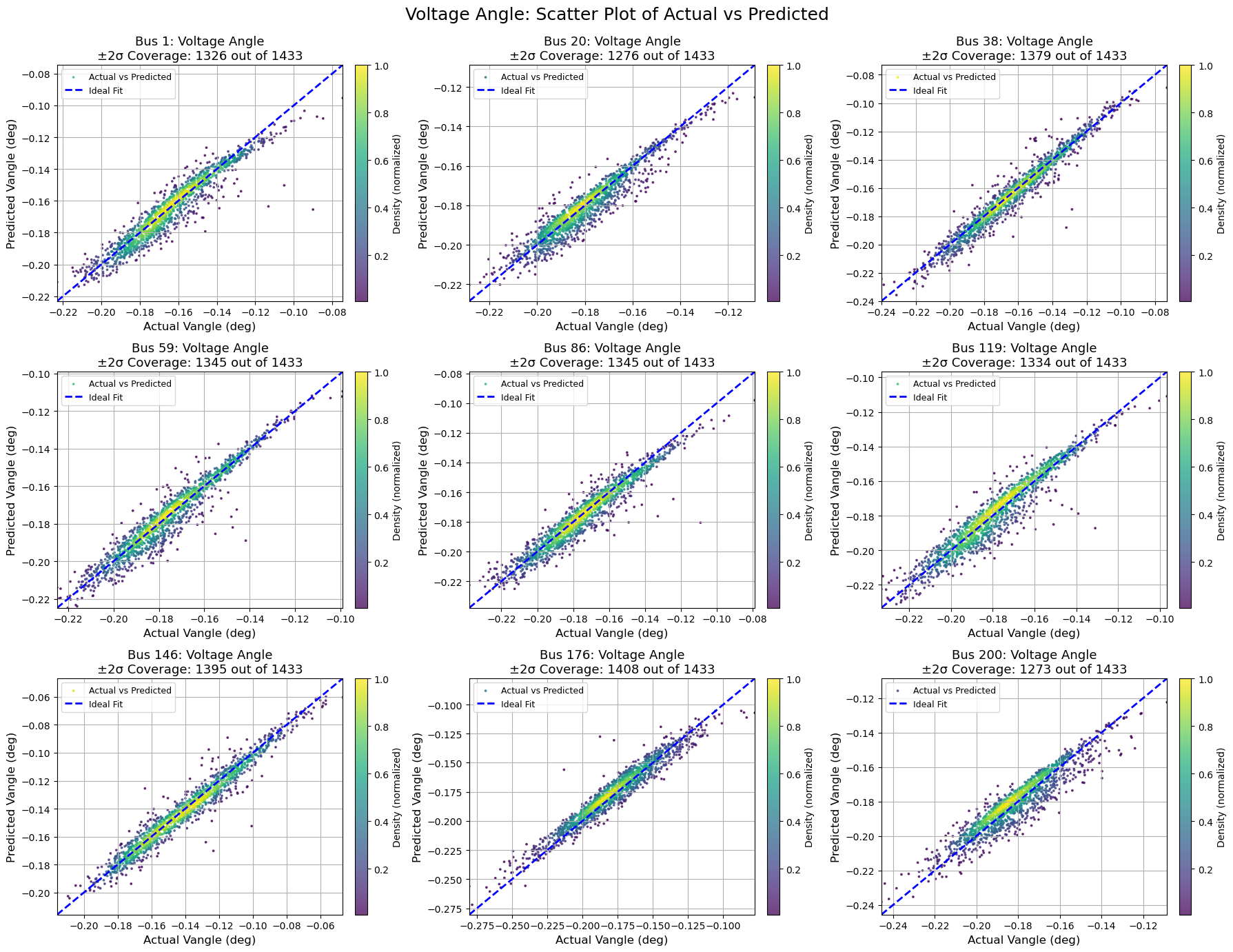}

    \par\vspace{1mm}
    {\scriptsize
    \makebox[0.48\textwidth]{\centering (a) Train}
    \hspace{1mm}
    \makebox[0.48\textwidth]{\centering (b) Test}
    }

    \caption{Parity plots from Stage 2 comparing actual and predicted results of voltage angle for the selected buses in the synthetic 200-bus system with AC-OPF, shown for the training data (left) and testing data (right)}
    \label{fig:train_test}
\end{figure*}

\begin{figure*}[!t]
    \centering
    \includegraphics[width=0.48\textwidth]{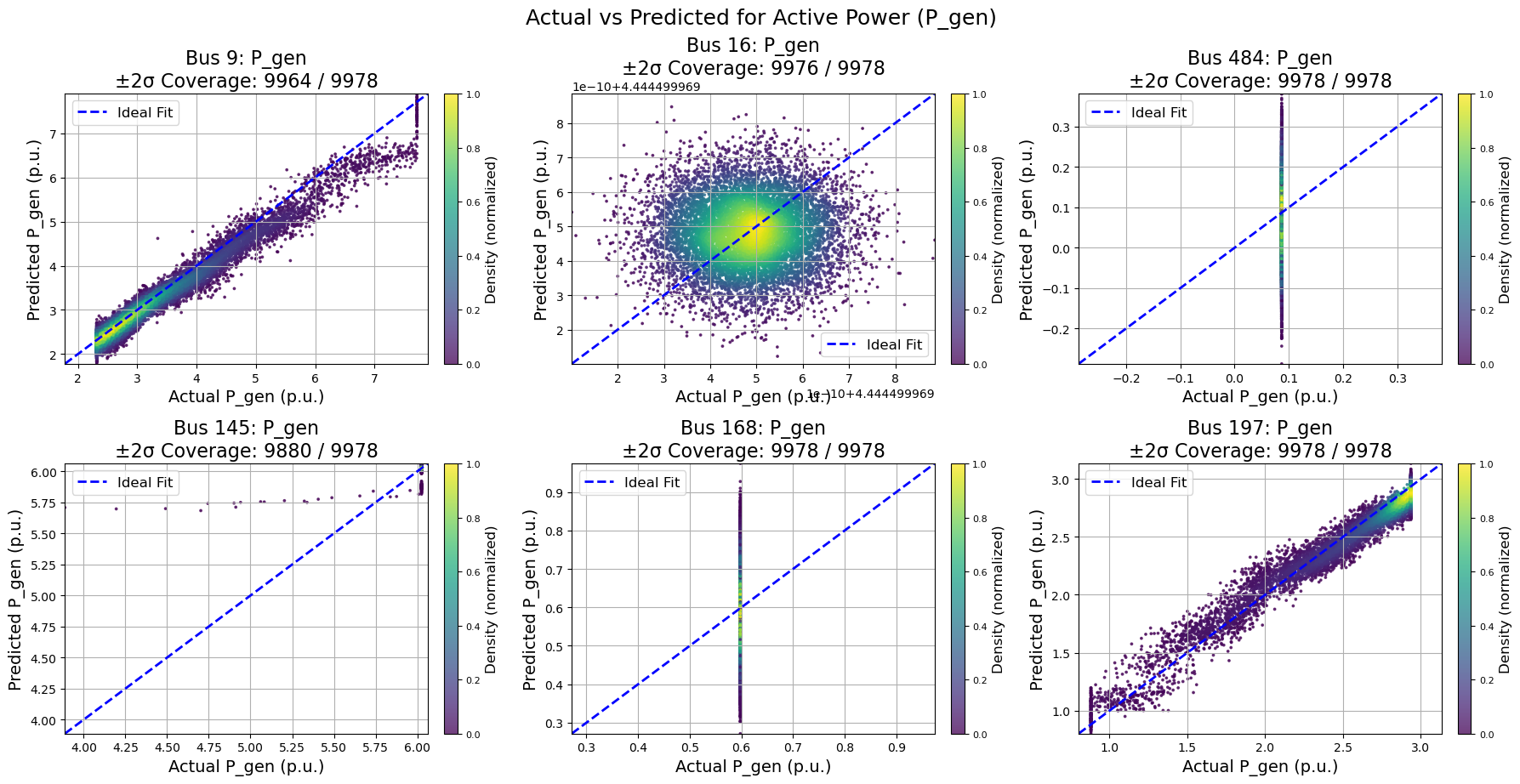}
    \hspace{1mm}
    \includegraphics[width=0.48\textwidth]{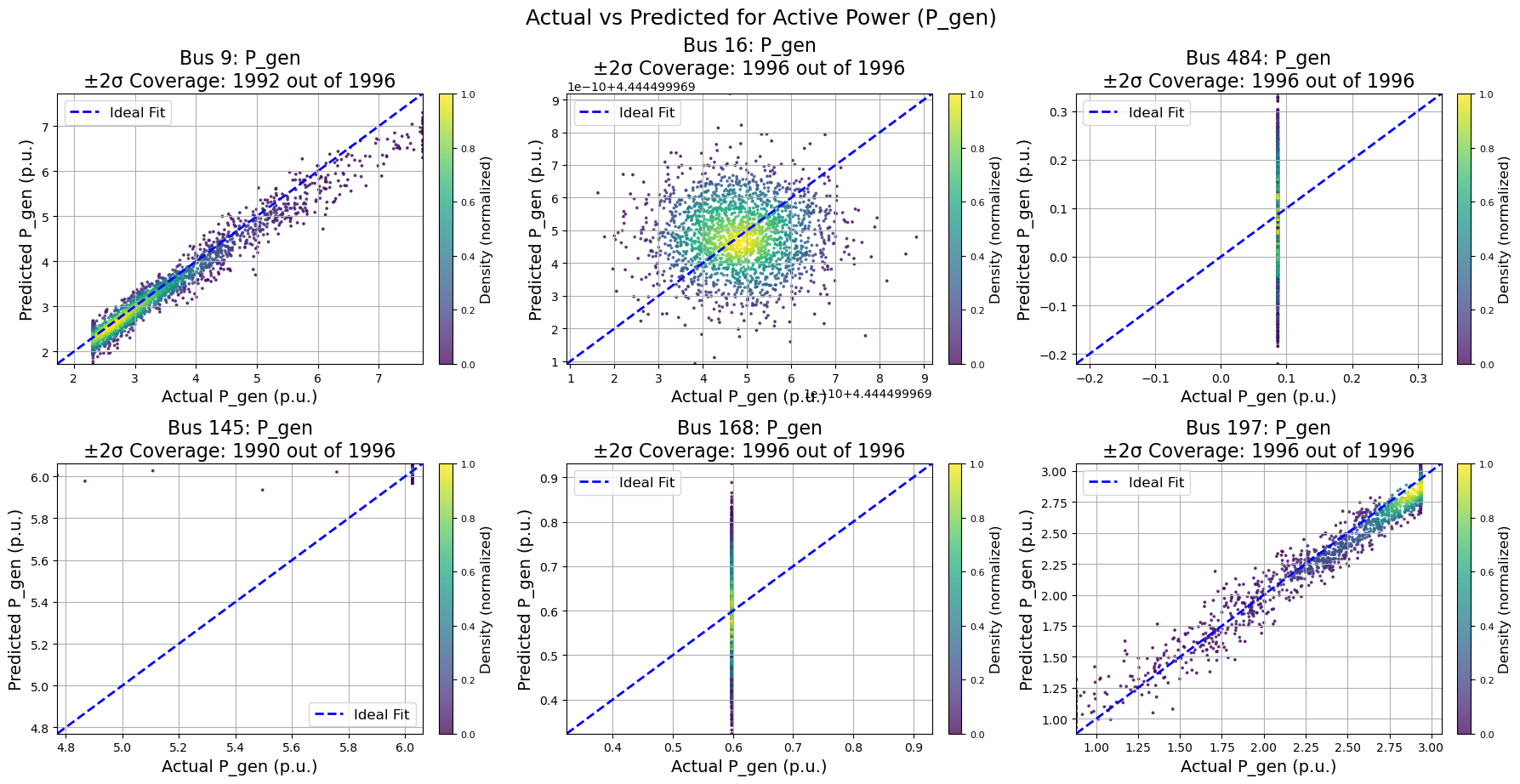}

    \par\vspace{1mm}
    {\scriptsize
    \makebox[0.48\textwidth]{\centering (a) Train}
    \hspace{1mm}
    \makebox[0.48\textwidth]{\centering (b) Test}
    }

    \caption{Parity plots from Stage 1 comparing actual and predicted active power set points for the selected buses in the synthetic 500-bus system for AC-OPF, shown for the training data (left) and testing data (right)}
    \label{fig:train_test}
\end{figure*}

\begin{figure*}[!t]
    \centering
    \includegraphics[width=0.48\textwidth]{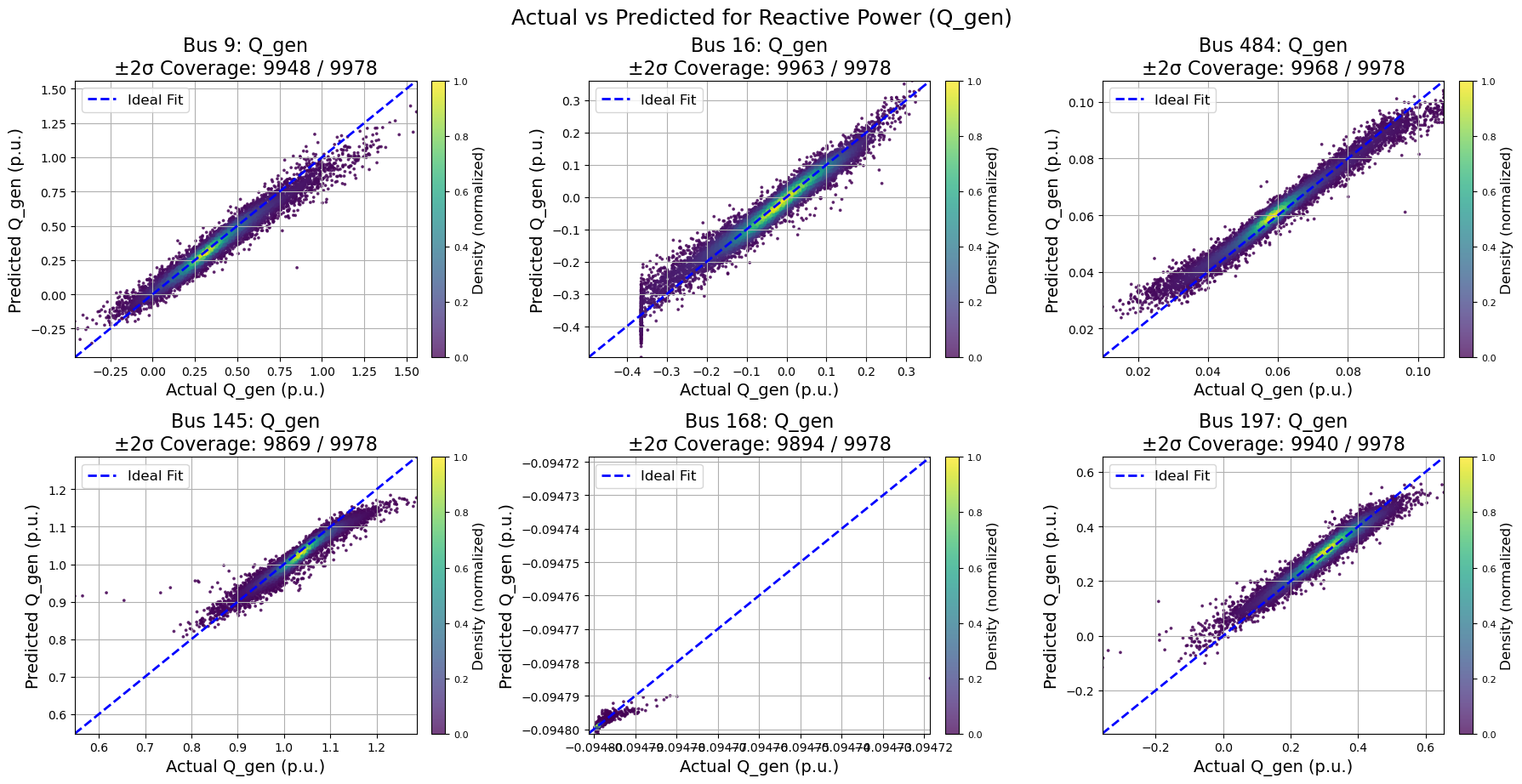}
    \hspace{1mm}
    \includegraphics[width=0.48\textwidth]{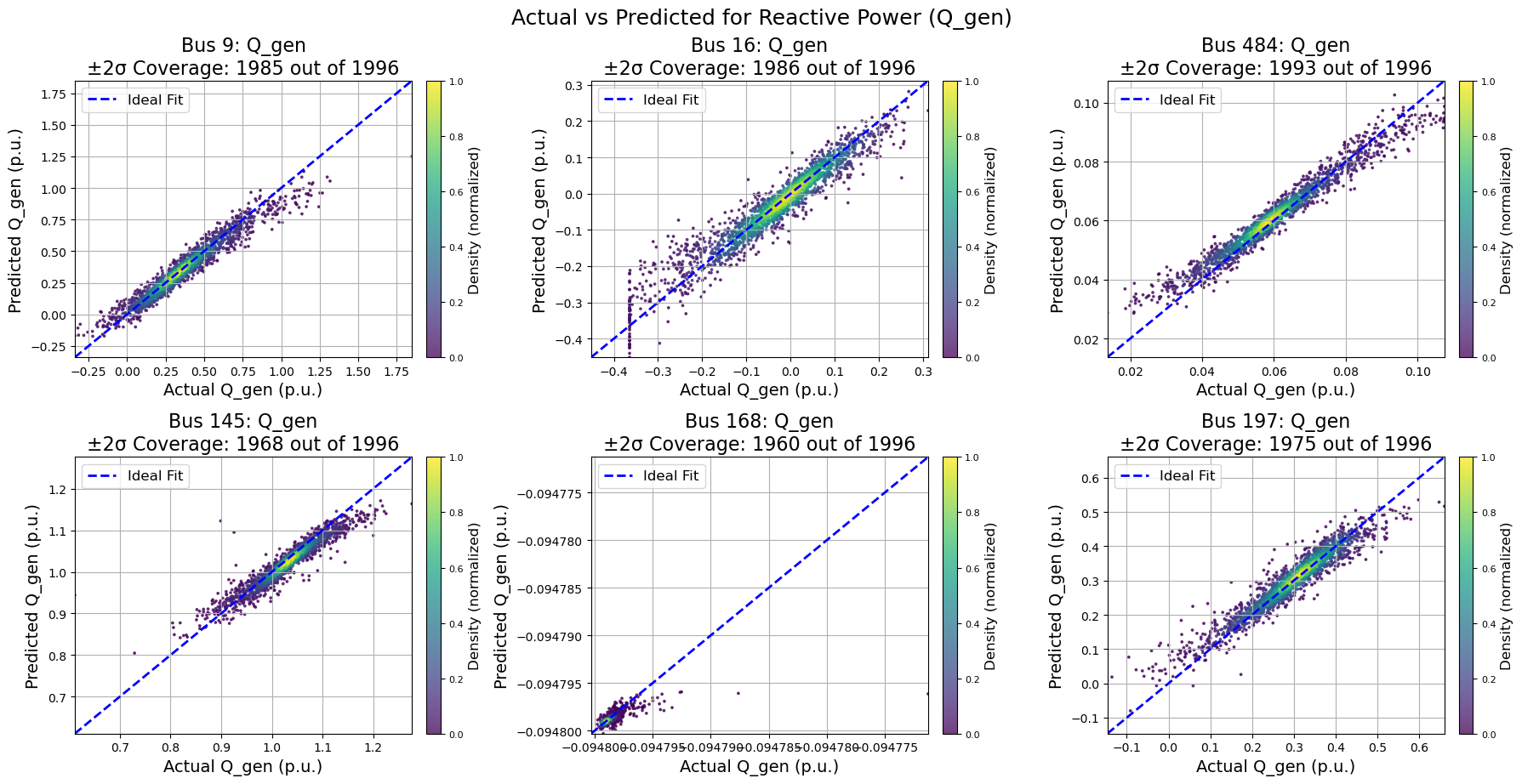}

    \par\vspace{1mm}
    {\scriptsize
    \makebox[0.48\textwidth]{\centering (a) Train}
    \hspace{1mm}
    \makebox[0.48\textwidth]{\centering (b) Test}
    }

    \caption{Parity plots from Stage 1 comparing actual and predicted active power set points for the selected buses in the synthetic 500-bus system for AC-OPF, shown for the training data (left) and testing data (right)}
    \label{fig:train_test}
\end{figure*}

\begin{figure*}[!t]
    \centering
    \includegraphics[width=0.48\textwidth]{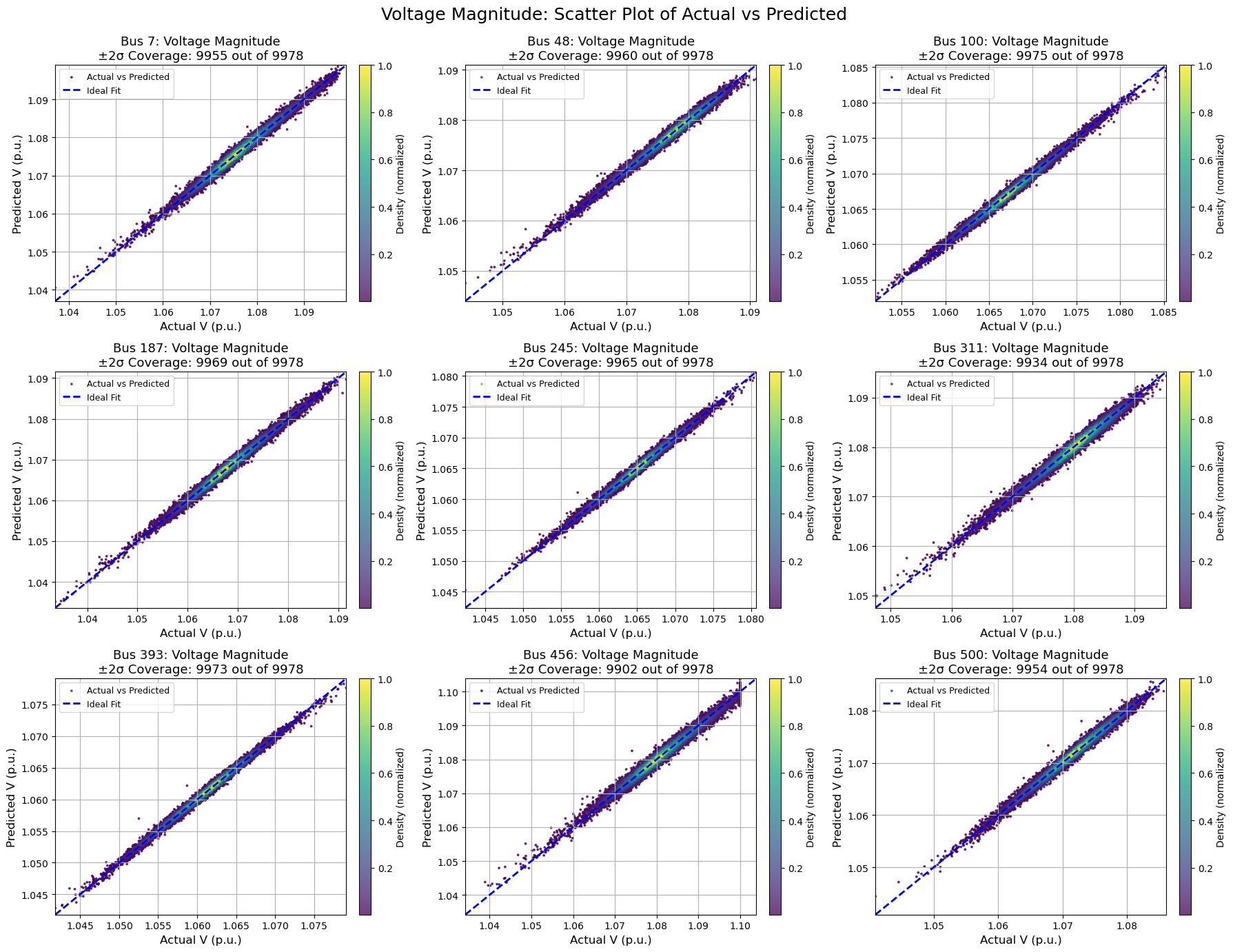}
    \hspace{1mm}
    \includegraphics[width=0.48\textwidth]{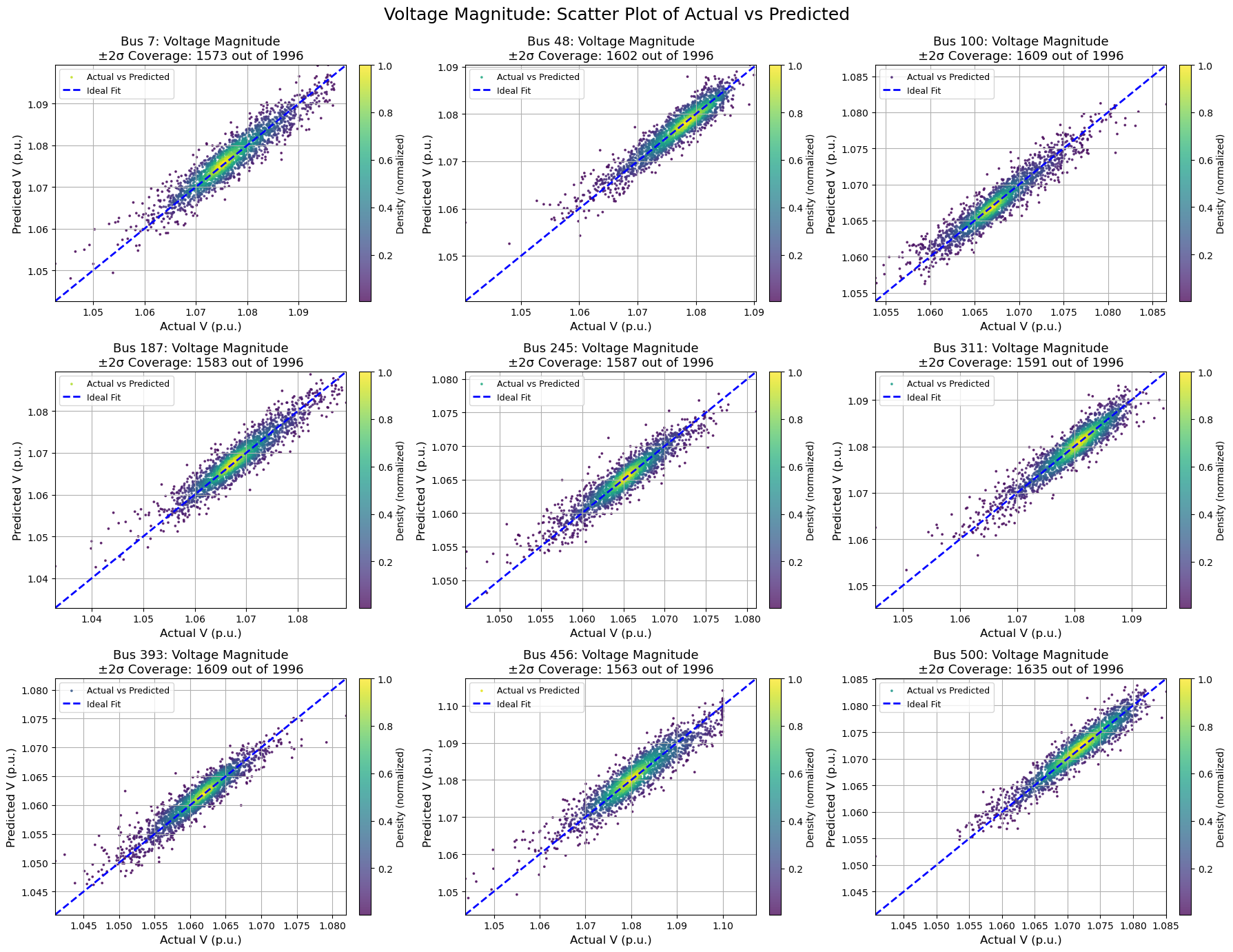}

    \par\vspace{1mm}
    {\scriptsize
    \makebox[0.48\textwidth]{\centering (a) Train}
    \hspace{1mm}
    \makebox[0.48\textwidth]{\centering (b) Test}
    }

    \caption{Parity plots from Stage 2 comparing actual and predicted results of voltage magnitude for the selected buses in the synthetic 500-bus system with AC-OPF, shown for the training data (left) and testing data (right)}
    \label{fig:train_test}
\end{figure*}

\begin{figure*}[!t]
    \centering
    \includegraphics[width=0.48\textwidth]{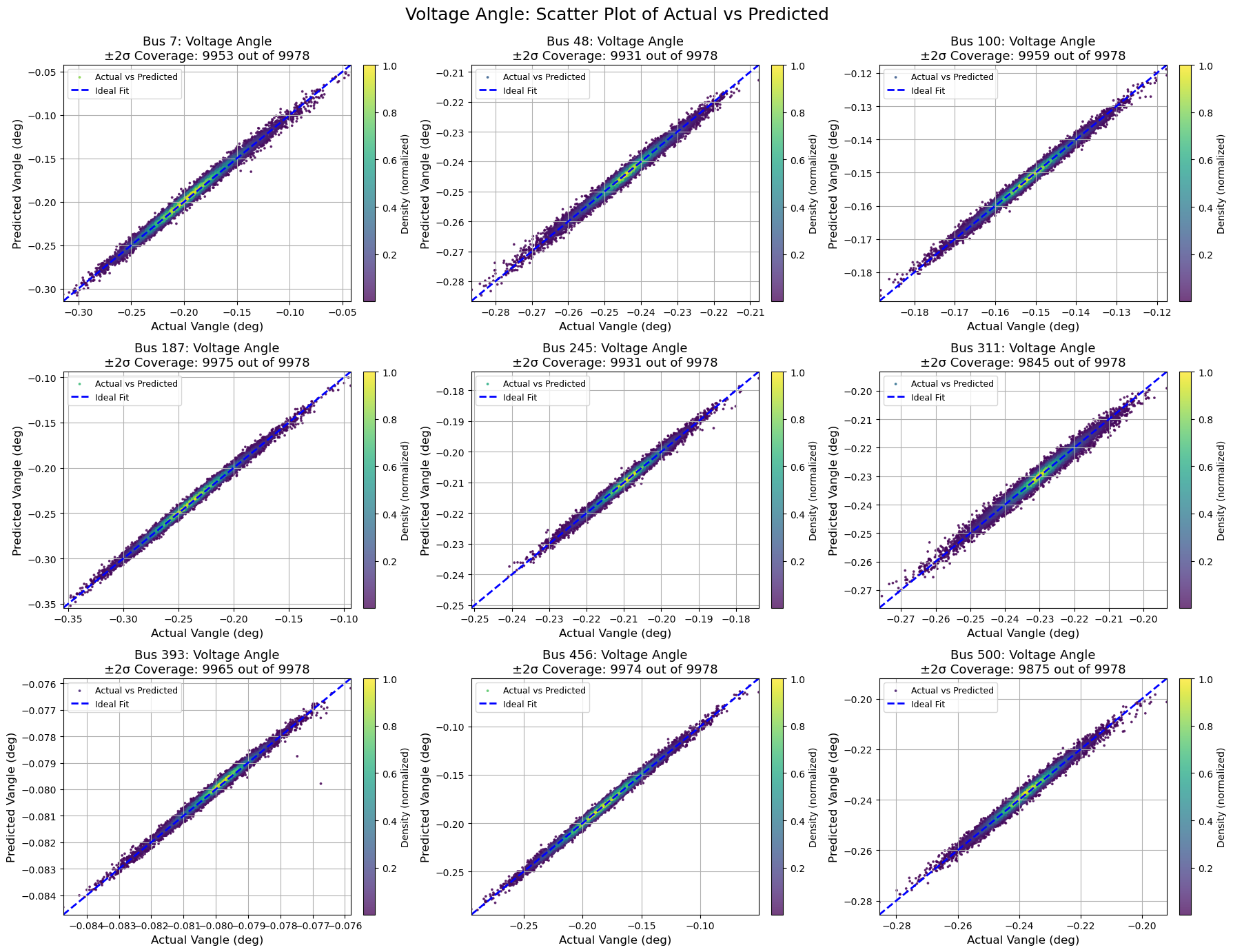}
    \hspace{1mm}
    \includegraphics[width=0.48\textwidth]{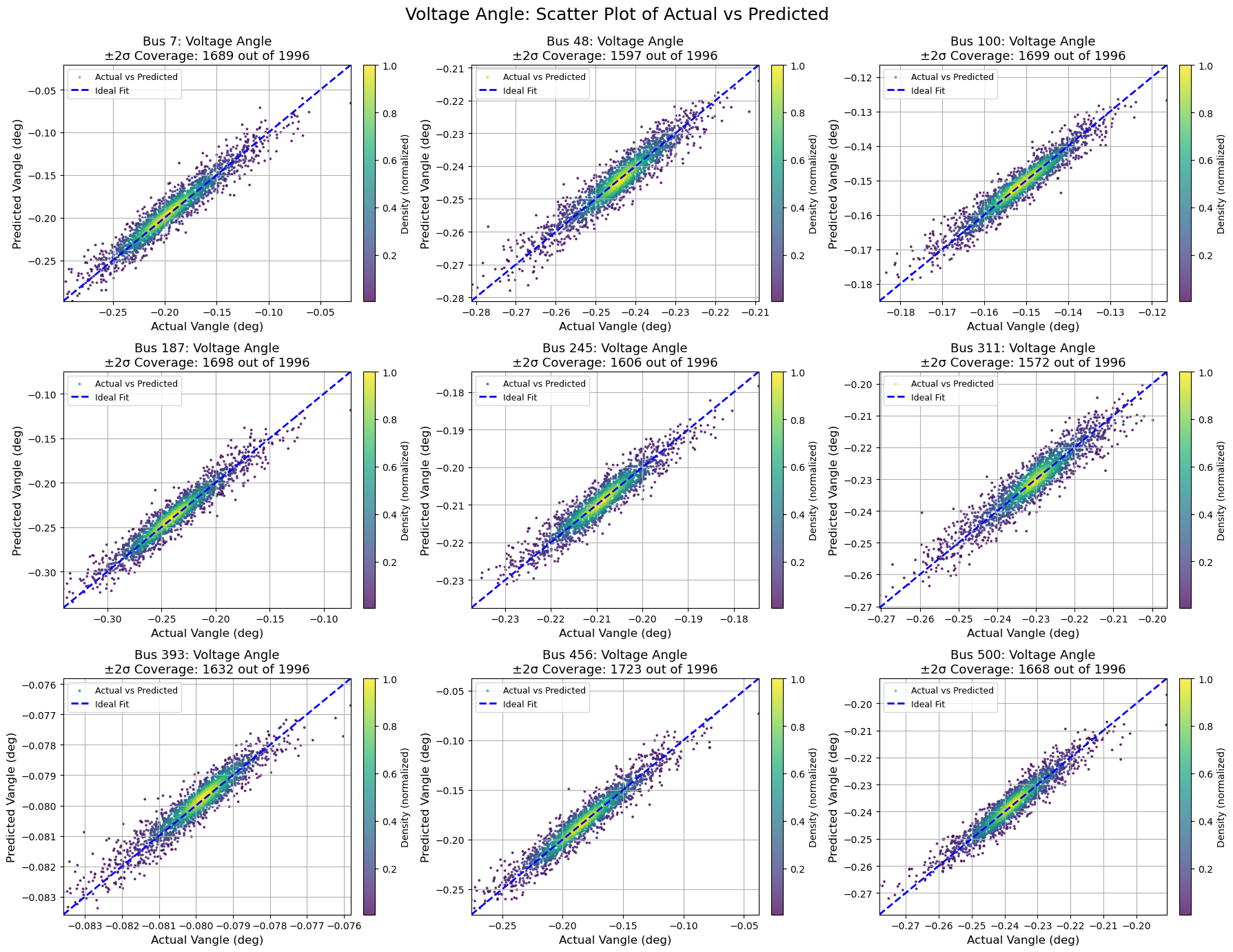}

    \par\vspace{1mm}
    {\scriptsize
    \makebox[0.48\textwidth]{\centering (a) Train}
    \hspace{1mm}
    \makebox[0.48\textwidth]{\centering (b) Test}
    }

    \caption{Parity plots from Stage 2 comparing actual and predicted results of voltage angle for the selected buses in the synthetic 500-bus system with AC-OPF, shown for the training data (left) and testing data (right)}
    \label{fig:train_test}
\end{figure*}

\section{Conclusion}

This work introduced PowerModels-ACOPF-AI, a scalable two-stage Bayesian surrogate that predicted AC-OPF set points and steady-state voltages/angles with calibrated uncertainty and demonstrated the ability to self-improve when faced with novel operating regimes. Trained on AC-OPF solutions generated via PowerModels.jl, the model generalized from a 30-bus benchmark to 200- and 500-bus systems and maintained high empirical coverage (\(\approx 95\text{--}99\%\) within \(\pm 2\sigma\)), even under substantial variability in load and nondispatchable renewables. When deliberately stressed with out-of-distribution inputs, its performance-aware on-the-fly learning mechanism targeted data generation and retraining, thereby restoring coverage to \(\geq 97\text{--}100\%\) and demonstrating practical robustness for evolving grids.

Beyond predictive accuracy, the model’s ability to deliver real-time predictions along with reliable uncertainty estimates made it valuable in two main roles: (i) as a real-time advisor for operators seeking fast, physically consistent AC guidance with credible intervals and (ii) as a warm start generator that reduced solve times for conventional nonlinear programs. Current limitations included the absence of explicit feasibility guarantees and contingency/security constraints within the surrogate, and we did not report end-to-end wall-clock comparisons against state-of-the-art solvers in control-room conditions. Additionally, the BNN model, like other ML-based approaches, cannot guarantee global optimality, particularly for complex, non-convex problems such as AC-OPF, and is inherently subject to prediction uncertainty\cite{b23}. The model is trained on locally optimal solutions and therefore primarily ensures proximity to local optima. Nevertheless, the predicted solutions are typically near-optimal and can be effectively used as warm-start initializations to accelerate the convergence of conventional optimization methods. Future work will integrate physics-informed feasibility layers and constraint screening, extend the framework to security-constrained and stochastic/robust OPF, incorporate topology changes and N-1/N-1-1 contingencies, and couple the surrogate with probabilistic weather-driven renewable forecasts.

\newpage

\EOD


\clearpage
\begin{thebibliography}{00}
\bibitem{b1} M. B. Cain, R. P. O’neill, A. Castillo , ``History of optimal power flow and formulations,'' Federal Energy Regulatory Commission, 2012, pp. 1-36.

\bibitem{b2} N.S. Rau, ``Issues in the path toward an RTO and standard markets,'' \emph{IEEE Trans. on Power Systems}, vol. 18, no. 2, pp.  435–443, May 2003a, doi: 10.1109/TPWRS.2003.810709.

\bibitem{b3} B. Stott and E. Hobson, "Power System Security Control Calculations Using Linear Programming, Part II," \emph{IEEE Trans. on Power Apparatus and Systems}, vol. PAS-97, no. 5, pp. 1721-1731, Sept. 1978, doi: 10.1109/TPAS.1978.354665.

\bibitem{b4} H. W. Dommel and W. F. Tinney, "Optimal Power Flow Solutions," \emph{IEEE Trans. on Power Apparatus and Systems}, vol. PAS-87, no. 10, pp. 1866-1876, Oct. 1968, doi: 10.1109/TPAS.1968.292150.

\bibitem{b5} S. J. Wang, S. M. Shahidehpour, D. S. Kirschen, S. Mokhtari and G. D. Irisarri, "Short-term generation scheduling with transmission and environmental constraints using an augmented Lagrangian relaxation," \emph{IEEE Trans. on Power Systems}, vol. 10, no. 3, pp. 1294-1301, Aug. 1995, doi: 10.1109/59.466524.

\bibitem{b6} R. A. Jabr, A. H. Coonick and B. J. Cory, "A homogeneous linear programming algorithm for the security constrained economic dispatch problem," \emph{IEEE Trans. on Power Systems}, vol. 15, no. 3, pp. 930-936, Aug. 2000, doi: 10.1109/59.871715.

\bibitem{b7} F. Bouffard, F. D. Galiana and A. J. Conejo, "Market-clearing with stochastic security-part I: formulation," \emph{IEEE Trans. on Power Systems}, vol. 20, no. 4, pp. 1818-1826, Nov. 2005, doi: 10.1109/TPWRS.2005.857016.

\bibitem{b8} H. Singh, S. Hao and A. Papalexopoulos, "Transmission congestion management in competitive electricity markets," \emph{IEEE Trans. on Power Systems}, vol. 13, no. 2, pp. 672-680, May 1998, doi: 10.1109/59.667399.

\bibitem{b9} A. S. Zamzam and K. Baker, "Learning Optimal Solutions for Extremely Fast AC Optimal Power Flow," \emph{2020 IEEE Int. Conf. on Communications, Control, and Computing Technologies for Smart Grids (SmartGridComm)}, Tempe, AZ, USA, 2020, pp. 1-6, doi: 10.1109/SmartGridComm47815.2020.9303008.


\bibitem{b10} F. Hasan, A. Kargarian and A. Mohammadi, "A Survey on Applications of Machine Learning for Optimal Power Flow," \emph{2020 IEEE Texas Power and Energy Conference (TPEC)}, College Station, TX, USA, 2020, pp. 1-6, doi: 10.1109/TPEC48276.2020.9042547

\bibitem{b11} B. Jiang, Q. Wang, S. Wu, Y. Wang, G. Lu,"Advancements and Future Directions in the Application of Machine Learning to AC Optimal Power Flow: A Critical Review," \emph{Energies} vol. 17, pp. 1381, 2024, doi.org/10.3390/en17061381

\bibitem{b12} H. Khaloje, M. Dolányi, J. F Toubeau, F. Vallée, “Review of machine learning techniques for optimal power flow”, \emph{Applied Energy}, vol. 388, pp. 125637, 2025, doi.org/10.1016/j.apenergy.2025.125637.

\bibitem{b13} C. Coffrin, R. Bent, K. Sundar, Y. Ng and M. Lubin, "PowerModels. JL: An Open-Source Framework for Exploring Power Flow Formulations," \emph{2018 Power Systems Computation Conference (PSCC)}, Dublin, Ireland, 2018, pp. 1-8, doi: 10.23919/PSCC.2018.8442948.

\bibitem{b14} K. Baker, "Emulating AC OPF Solvers With Neural Networks," \emph{IEEE Trans. on Power Systems}, vol. 37, no. 6, pp. 4950-4953, Nov. 2022, doi: 10.1109/TPWRS.2022.3195097

\bibitem{b15} W. A. Astudillo, F. Astudillo-Salinas and S. P. Torres, "Evaluation of a Machine Learning-based Algorithm for AC Optimal Power Flow," \emph{2024 IEEE Eighth Ecuador Technical Chapters Meeting (ETCM)}, Cuenca, Ecuador, 2024, pp. 1-6, doi: 10.1109/ETCM63562.2024.10746103.

\bibitem{b16} Y. Zhou et al., "A Data-driven Method for Fast AC Optimal Power Flow Solutions via Deep Reinforcement Learning," \emph{Journal of Modern Power Systems and Clean Energy}, vol. 8, no. 6, pp. 1128-1139, November 2020, doi: 10.35833/MPCE.2020.000522.

\bibitem{b17} W. Huang, M. Chen and S. H. Low, "Unsupervised Learning for Solving AC Optimal Power Flows: Design, Analysis, and Experiment," \emph{IEEE Trans. on Power Systems}, vol. 39, no. 6, pp. 7102-7114, Nov. 2024, doi: 10.1109/TPWRS.2024.3373399.

\bibitem{b18} M. Chatzos, T. W. K. Mak and P. V. Hentenryck, "Spatial Network Decomposition for Fast and Scalable AC-OPF Learning," \emph{IEEE Trans. on Power Systems}, vol. 37, no. 4, pp. 2601-2612, July 2022, doi: 10.1109/TPWRS.2021.3124726.

\bibitem{b19} K. Chen, S. Bose and Y. Zhang, "Physics-Informed Gradient Estimation for Accelerating Deep Learning-Based AC-OPF," \emph{IEEE Transactions on Industrial Informatics}, doi: 10.1109/TII.2025.3545080.

\bibitem{b24} Ugwumadu,C.;Tabarez,J.;
Drabold, D.A.; Pandey,A., "PowerModel-AI: A First On-the-Fly Machine Learning Predictor for AC Power FlowSolution," \emph{Energies}, Volume 18, Article 1968, April 2025,
18, 1968. https://doi.org/10.3390/
en18081968

\bibitem{b20} A.B. Birchfield, T. Xu, K.M. Gegner, K.S. Shetye, and T.J. Overbye, "Grid Structural Characteristics as Validation 
Criteria for Synthetic Networks," to appear, \emph{IEEE Transactions on Power Systems}, 2017.

\bibitem{b21} L. V. Jospin, H. Laga, F. Boussaid, W. Buntine, and M. Bennamoun, "Hands-On Bayesian Neural Networks—A Tutorial for Deep Learning Users," \emph{IEEE Computational Intelligence Magazine}, vol. 17, no. 2, pp. 29-48, May 2022, doi: 10.1109/MCI.2022.3155327.

\bibitem{b22} Reactive Power Capability and Interconnection Requirements for PV and Wind Plants, Sandia National Laboratories. Available online: \url{http://www.esig.energy/wiki-main-page/reactive-power-capability-and-interconnection-requirements-for-pv-andwind-plants/#cite_note-1}.

\bibitem{b23} Fioretto, F., Mak, T. W., Van Hentenryck, P., "Predicting ac optimal power flows: Combining deep learning and lagrangian dual methods," \emph{In Proceedings of the AAAI conference on artificial intelligence}, Vol. 34, no. 01, pp. 630-637, April 2020.

\end{thebibliography}
\end{document}